\documentclass[preprint,authoryear,12pt]{elsarticle}

\usepackage{graphics}
\usepackage{float}
\usepackage{graphicx}
\usepackage{wrapfig}

\usepackage[draft]{todonotes}   % notes shown

\usepackage[labelfont=bf]{caption}
\usepackage{caption}
\usepackage{color}
\usepackage{amsmath} % Mathe
\usepackage{mathtools} % Mathe
\usepackage{amsfonts} % Mathesymbole
\usepackage{amssymb}
\usepackage{array,longtable}

\usepackage{array}
\usepackage{booktabs}
\usepackage{rotating}
\usepackage{multirow}
\usepackage{multicol}
\usepackage{lscape}
\usepackage{subfig}

\usepackage[utf8]{inputenc}
\usepackage[T1]{fontenc}
\usepackage{textcomp}
\usepackage{gensymb}

\usepackage[export]{adjustbox}

\usepackage{algorithm,algorithmicx,algpseudocode}

\usepackage{color}

\usepackage[colorlinks=true,linkcolor=blue,citecolor=blue]{hyperref}

\usepackage[draft]{todonotes}   % notes shown

\usepackage{soul}

\definecolor{darkgreen}{rgb}{0.53, 0.66, 0.42}
\definecolor{azure}{rgb}{0.0, 0.5, 1.0}

\journal{Medical Image Analysis}

\begin{document}

\begin{frontmatter}

%% Title, authors and addresses

%% use the tnoteref command within \title for footnotes;
%% use the tnotetext command for the associated footnote;
%% use the fnref command within \author or \address for footnotes;
%% use the fntext command for the associated footnote;
%% use the corref command within \author for corresponding author footnotes;
%% use the cortext command for the associated footnote;
%% use the ead command for the email address,
%% and the form \ead[url] for the home page:
%%
%% \title{Title\tnoteref{label1}}
%% \tnotetext[label1]{}
%% \author{Name\corref{cor1}\fnref{label2}}
%% \ead{email address}
%% \ead[url]{home page}
%% \fntext[label2]{}
%% \cortext[cor1]{}
%% \address{Address\fnref{label3}}
%% \fntext[label3]{}

%Multi-view Multi-layer
\title{MGN-Net: a multi-view graph normalizer for integrating heterogeneous biological network populations}

\author[BASIRA]{Mustafa Burak G\"{u}rb\"{u}z}
\author[BASIRA,DUNDEE]{Islem Rekik\corref{cor}}
\author{and for the Alzheimer's Disease Neuroimaging Initiative}
\cortext[cor]{Corresponding author: irekik@itu.edu.tr; \url{http://basira-lab.com/}. Data used in preparation of this article were obtained from the Alzheimer's Disease Neuroimaging Initiative (ADNI) database (\url{adni.loni.usc.edu}). As such, the investigators within the ADNI contributed to the design and implementation of ADNI and/or provided data but did not participate in analysis or writing of this report. A complete listing of ADNI investigators can be found at: http://adni.loni.usc.edu/wp- content/uploads/how to apply/ADNI Acknowledgement List.pdf}
%\cortext[ADNI]{ }

\address[BASIRA]{BASIRA lab, Faculty of Computer and Informatics Engineering, Istanbul Technical University, Istanbul, Turkey}
\address[DUNDEE]{School of Science and Engineering, Computing, University of Dundee, UK \ }

%% use optional labels to link authors explicitly to addresses:
%% \author[label1,label2]{<author name>}
%% \address[label1]{<address>}
%% \address[label2]{<address>}

\begin{abstract}

With the recent technological advances, biological datasets, often represented by networks (i.e., graphs) of interacting entities, proliferate with unprecedented complexity and heterogeneity. Although modern network science opens new frontiers of analyzing connectivity patterns in such datasets, we still lack data-driven methods for extracting an integral connectional fingerprint of a multi-view graph population, let alone disentangling the typical from the atypical variations across the population samples. We present the multi-view graph normalizer network (MGN-Net\footnote{\url{https://github.com/basiralab/MGN-Net}}), a graph neural network based method to normalize and integrate a set of multi-view biological networks into a single connectional template that is centered, representative, and topologically sound. We demonstrate the use of MGN-Net by discovering the connectional fingerprints of healthy and neurologically disordered brain network populations including Alzheimer's disease and Autism spectrum disorder patients. Additionally, by comparing the learned templates of healthy and disordered populations, we show that MGN-Net significantly outperforms conventional network integration methods across extensive experiments in terms of producing the most centered templates, recapitulating unique traits of populations, and preserving the complex topology of biological networks. Our evaluations showed that MGN-Net is powerfully generic and easily adaptable in design to different graph-based problems such as identification of relevant connections, normalization and integration.

\end{abstract}

\begin{keyword}
%% keywords here, in the form: keyword \sep keyword
connectional brain templates \sep multi-view graph normalizer network \sep population multiview brain network integration \sep graph convolution networks 
%% MSC codes here, in the form: \MSC code \sep code
%% or \MSC[2008] code \sep code (2000 is the default)

\end{keyword}

\end{frontmatter}

%\linenumbers

%% ***************************************************************************** %%
\section{Introduction}
%% ***************************************************************************** %%

\begin{figure}[ht!]
	\centering
	\includegraphics[width=13cm]{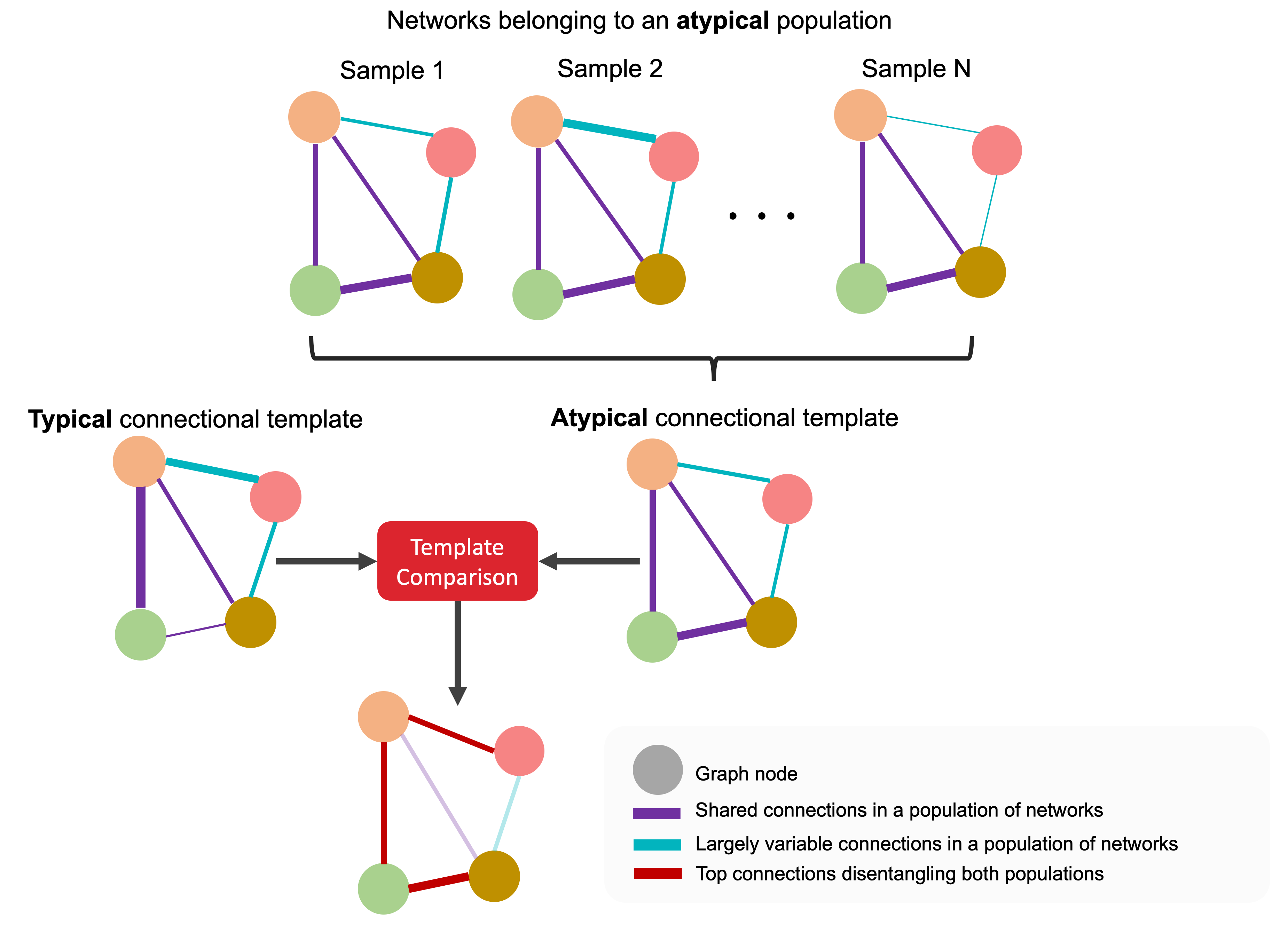}
	\caption{\emph{Connectional template comparison for identifying discriminative connections differentiating between typical and atypical populations of networks.}}
	\label{fig:1}
\end{figure}

Modern network science has introduced exciting new opportunities for understanding the underpinning mechanisms of biological systems by examining interactions within their components \citep{SystemsBio}. In the face of the ongoing `tsunami' of biological data collection spanning the range from genetic \citep{HGP} and metabolic networks \citep{KEGG} all the way up to social and economic systems \citep{contactNetwork}, data-driven network representations have allowed us to map complex interplay between components of biological systems such as genetic data by revealing gene co-expression and connectomic data by investigating correlations in neural signaling between different brain regions. Namely, graphs present a natural tool to study such interactions (or connections) in complex biological data including protein-protein interactions  \citep{PPI}, metabolic networks \citep{metabolic}, and brain connectivity networks which span the field of network neuroscience \citep{CBN}. In the former context, network neuroscience proposes to encode the brain wiring in a graph by representing the brain regions as nodes and their interactions as edges linking those nodes, which has propelled the development of advanced network-based analysis techniques of the brain construct \citep{NN}. Particularly given the recent proliferation in large and multi-modal connectomic datasets such as the Human Connectome Project \citep{HCP} acquired using multiple neuroimaging modalities including structural T1-weighted, diffusion, and functional magnetic resonance imaging (MRI), it is not always obvious how
to integrate multi-modal connectomic data together \citep{van2016human}, nor easy to do so in practice, in order to first understand how the brain's structural, morphological and functional levels interlink to form this integrated complex system, and then identify typical and atypical connectional trends fingerprinting the human brain. This is substantially due to the large variability in brain connectivity across individuals, which limits our ability to disentangle the `healthy' brain connectional variability from the `pathological' variability. For instance, two individuals who largely differ in particular brain connections might not indicate that one of them has a pathological connection, this difference can still fit healthy connectional brain patterns. To distinguish between healthy and disordered connectional variability, we need to define a `normalization' or `standardization' process of brain networks  (\textbf{Fig.}~\ref{fig:1}). Eventually, we hypothesize that reducing inter-subject variability in both healthy and disordered populations through a \emph{normalization process} will contribute towards helping better identify `pathological' alterations in brain networks as deviations from the `standard/normalized' brain network representation. Without any loss of generalizability, this line of reasoning extends to other biological networks with multiple views, meaning that each sample is represented by a set of networks, where each network view captures unique traits of the sample (for example, functional, morphological and structural).

Here we propose a novel multi-view graph integration (MGI) method which produces a unified normalized connectional representation of a population of multi-view  networks; for example a connectional brain template (CBT) from a heterogeneous population of multi-view brain networks. Specifically, our MGI method taps into the power of deep-learning multi-disciplinary architectures, which can handle large-scale, highly non-linear and heterogeneous datasets. Although deep learning has been recently used to regress brain networks \citep{bronstein2017geometric}, there has not been any work on applying deep learning for fusing networks in general. Critically, fusing multi-view biological networks is an uncharted territory where network science and deep learning have not cross-fertilized, especially the emerging field of geometric deep learning on graphs and manifolds (i.e., graph neural networks). Here we set out to integrate a population of multi-view biological graphs with the aim to estimate a representative reference connectional template by normalizing connections across the population samples, which is an essential step for group comparison studies as well as discovering the integral signature of an anomaly in a given population (e.g., disordered) by comparison with a typical connectional template (e.g., healthy). In this context, we hypothesize that a population-driven connectional template satisfies the following properties: (1) well-centeredness, (2) discriminativeness, and (3) topological soundness. A \emph{well-centered} template occupies the `center' of a population by achieving the minimum distance to all population samples. A \emph{discriminative} template implies that the estimated template consistently captures the unique and distinctive traits of a population. Last but not least, convergent studies show that a large variety of biological networks has extraordinarily complex yet highly organized topological patterns such as the spatially economical layout of brain regions that are likely to be a consequence of the conservation of wiring costs being an important selection pressure on the evolution of brain networks \citep{CBN}. Therefore, the estimated template should be topologically sound by preserving the population topological properties during the normalization process. %\textcolor{caribbeangreen}{This suggests that observed brain organisation may be the consequence of a set of consistent principles that are expressed across multiple scales (gene expression, cytoarchitecture, cortical wiring and macroscale function).}                   

From a deep learning perspective, methods for multi-view network integration are currently lacking. The simplest way to integrate a set of biological networks is to linearly average them. However, such a normalization technique alone is very sensitive to outliers and cannot be generalized to blend information of multi-view network datasets with heterogeneous distributions. Currently, the prevailing technique for non-linear network integration is similarity network fusion (SNF) \citep{SNF}, which is based on message passing theory \citep{Message}. SNF aims to estimate a status matrix for each network that carries the whole information in the networks and a sparse local matrix that only takes up to top-$k$ neighbors into consideration. Next, an iterative integration step is conducted to update each status network through diffusing mean global structure of remaining networks and along with the sparse local network. Even though SNF is a powerful tool since it is a generic unsupervised technique, it comes with strong assumptions such as emphasizing the top $k$ local connections for each node and equally averaging the global topology of complementary networks for each iterative update to ultimately merge them. Another very recent approach, the netNorm \citep{netnorm}, utilizes a graph-based feature selection along with SNF to integrate multi-view networks. netNorm first constructs a high-order graph using cross-view connectional features as nodes and their Euclidean distance as a dissimilarity measure to select the most centered ones across the population. Next, for each network view, it composes a mosaic of the selected edges across subjects and eventually integrates the mosaic network views into a single network using SNF. Despite the promising results on multi-view datasets, netNorm has recognized limitations. First, it uses the Euclidean distance as a predefined metric for selecting the most representative connections which might fail to capture complex non-linear patterns in the given population. Second, netNorm consists of independent feature extraction, feature selection, and fusion steps. These fully independent steps cannot provide feedback to each other in order to globally optimize the template estimation process. Therefore, the pipeline is agnostic to cumulative errors. More importantly, both SNF \citep{SNF} and netNorm \citep{netnorm} do not have any mechanism to preserve complex topological patterns in biological networks during the integration process, which is undeniably substantial for outputting topologically sound connectional templates.

Here we propose the multi-view graph normalizer network (MGN-Net), a novel graph neural network (GNN) based method  for integrating and normalizing a set of multi-view graphs to \emph{learn} a representative connectional template for a given population. Our approach is inspired by cutting-edge, but so far neglected GNN frameworks \citep{kipf2016, FastLocalized, xu2019powerful, GAttention} in the field of network integration. It is also distinct in that it circumvents the need for handcrafted steps and general assumptions as it learns how to estimate the best template within an end-to-end optimization framework. GNNs are an emerging subfield of deep learning which extends the idea of convolutional neural networks (CNNs) \citep{CNNImage} to non-Euclidean data such as graphs and surfaces. GNNs achieved remarkable results in several recent biomedical data analysis studies such as disease classification \citep{Breast-Cancer-Subtype-Classification,ASD-and-AD} and protein interaction prediction \citep{protein-molecular-surfaces-using-gdl,protein-interface-prediction-usingGDL}. MGN-net capitalizes on graph neural network layers to explore implicit patterns that exist in the population of multi-view graphs and estimate the best template that is well-centered, discriminative, and topologically sound. 

This work presents an extension to the recent conference MICCAI 2020 paper \citep{DGN}. The method presented in \citep{DGN} introduces the Subject Normalization Loss (SNL) function for optimizing the proposed Deep Graph Normalizer (DGN) architecture. However, SNL does not constrain the integration process in terms of maintaining the complex topology of biological networks. Furthermore, it does not evaluate the topological soundness of generated connectional templates. To address these limitations, we further propose a novel loss function that penalizes the deviation from the ground-truth node strength \emph{topological} distribution. Through additional experiments, we show that this novel loss function not only enforces the learning of more topologically sound templates but also increases the performance in terms of centeredness and preserving discriminative traits of the network populations during the population multigraph integration. 

%\citep{DGN} measures the discriminativeness of generated templates by calculating the overlap rate between features selected by template comparison against those identified by independent supervised feature selection method. This type of evaluation assumes that the features selected by the independent method can be utilized as a ground truth.  Even though this type of evaluation is used in previous works \citep{netnorm}, it is a strong assumption to make.  Instead, in this work, we set up a different benchmark task where features selected by templates fed into a classifier and the accuracy of the classification used as a metric for the discriminativeness. Additionally, we further explore the generated templates such as centeredness of preliminary outputs (i.e. subject biased CBTs, see section \emph{4.2. Centeredness of Subject Biased Templates}) and discuss future directions such as exploiting dynamic brains networks to reveal the trajectory of neurological diseases and the possibility of collaborative training using deep learning approaches such as federated learning \citep{federated} and split neural networks \citep{split} (see section \emph{5. Discussion and Conclusion}).

\begin{figure}[H]
	\centering
	\includegraphics[width=14cm]{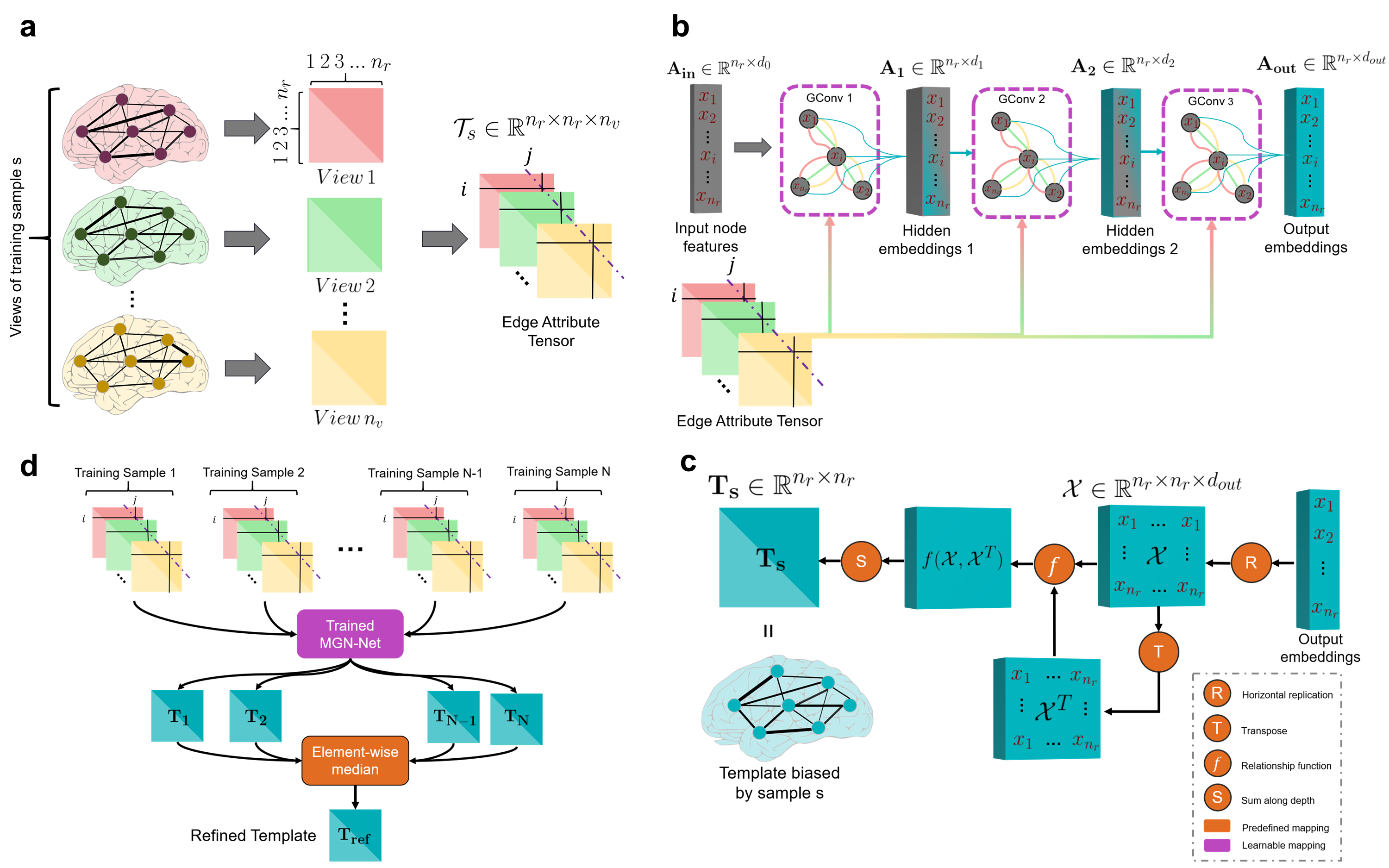}
	\caption{\emph{Overview of the proposed multi-view graph normalizer network (MGN-Net) architecture for integrating a population of multi-view graphs.} \textbf{a)} \textbf{Tensor representation of multi-view networks.}  We represent each input sample by a tensor $\mathcal{T}_{s} \in  \mathbb{R}^{n_{r} \times n_{r} \times n_{v}}$ where $n_r$ is the number of nodes and $n_v$ is the number of heterogeneous views. \textbf{b)} \textbf{Node embedding learning.} Abstract vector representation for each node of the input network is learned through 3 graph neural network layer. \textbf{c)} \textbf{Tensor representation of pair-wise relationship computation.} Calculation of pairwise relation of node embeddings is reformulated as a tensor operation for easy and efficient backpropagation. \textbf{d)} \textbf{Template refinement after training.}  We execute a post-training refinement step to eliminate any bias towards the given input subject by computing the median of all possible templates to create the ultimate template $\mathbf{T}_{ref}$ for the given population.
	}
	\label{fig:2}
\end{figure}

% %% ***************************************************************************** %%
\section{Method}
% %% ***************************************************************************** %%

\textbf{MGN-Net: a general framework to learn an integral and holistic connectional template of a multi-view graph population}. MGN-Net takes (\textbf{Fig.}~\ref{fig:2}\textcolor{blue}{--a}) two or more isomorphic weighted (possibly unweighted) multi-view graphs and maps them onto an output population center graph (i.e., connectional template). This learning task is fundamentally rooted in embedding connectivity patterns onto a high-dimensional vector representation for each node in each graph in the given population, namely a node embedding (\textbf{Fig.}~\ref{fig:2}\textcolor{blue}{--b}). During the embedding process, we preserve the unique domain-specific topological properties of the population graphs thanks to our novel topology-constrained normalization loss function which penalizes the deviation from ground-truth population topology. Next, we derive the template edges from the pairwise relationship of node embeddings (\textbf{Fig.}~\ref{fig:2}\textcolor{blue}{--c}). This relationship can be measured by any function that allows for deep network backpropagation depending on the application so that the MGN-Net can be trained in an end-to-end manner. Finally, we utilize our novel topology-constrained randomized loss function to minimize the distance between the population network views and the learned biological template.

Following the optimization of our MGN-Net architecture in the training stage, the learned connectional templates by the MGN-Net encapsulates both shared traits among samples and different yet complementary information offered by multi-view biological measurements. We used our method to integrate both small-scale and large-scale multi-view brain connectome datasets derived from T1-weighted MRI to estimate CBTs. We demonstrate that MGN-Net significantly outperforms other baseline network integration tools on both datasets in terms of centeredness (\textbf{Fig.}~\ref{fig:3}), preserving the highly organized topology of brain networks (\textbf{Table}~\ref{tab:1}), and identifying the most discriminative connections distinguishing between healthy and disordered brain states (\textbf{Table}~\ref{tab:2}). We tested the fingerprinting property of the learned CBTs by conducting a simple comparison between the learned healthy and disordered population CBTs and selecting the most discriminative connectional features that boost the classification accuracy of an independent machine learning-based diagnostic model. Our experiments highlighted the necessity of multi-view network integration to provide a normalized and expressive characterization of a population of brain multi-view graphs in both Alzheimer's diseases (AD) and the autism spectrum disorder (ASD) populations (\textbf{Table}~\ref{tab:3}).

\textbf{MGN-Net is an end-to-end graph neural network based architecture for normalizing multi-view graphs}.  To fully exploit the topological information offered by complex graph structures, we tap into the nascent field of GNNs on non-Euclidean spaces such as graphs instead of conventional learning methods where the implicit patterns are overlooked since graphs are projected onto a Euclidean space and processed similarly to any other numerical data. Given a population of multi-view samples, where each sample is represented by two or more graphs derived from different measurements (\textbf{Fig.}~\ref{fig:2}\textcolor{blue}{--a}), MGN-Net first feeds each multi-view sample through 3 graph convolutional layers. From layer to layer, deeper holistic embeddings are learned for each node recapitulating the complementary information offered by heterogeneous measurements (\textbf{Fig.}~\ref{fig:2}\textcolor{blue}{--b}). Next, we produce the normalized connectional template graph using pairwise relationships of node embeddings outputted by the final layer (\textbf{Fig.}~\ref{fig:2}\textcolor{blue}{--c}).

To evaluate the representativeness of the learned population templates, we propose the topology-constrained normalization loss (TCNL) function. The TCNL loss is composed of two parts. First, we compute the mean Frobenius distance between the learned template and a random subset of training samples to evaluate its centredness. Second, we compute the Kullback-Leibler divergence of node strength distributions of the connectional template and the random subset of training samples with the aim of measuring deviation from the real topology of the population multi-view graphs. We have particularly chosen the node strength distribution to constrain our loss optimization given that node strength presents a simple yet the most fundamental network measure and other advanced topological measures such as centrality measures depend on it \citep{CBN}. Note that TCNL compares the learned graph template with other multi-view graphs in the population, therefore the MGN-Net, by learning the most essential connectivities and normalizing trivial variabilities specific to samples, maps each multi-view graph to a population-representative template. The randomization of the training samples used for network normalization induces a loss regularization effect since it is much easier for the model to overfit if the loss is calculated against the whole dataset in each iteration. Moreover, the random sampling size is predefined as a hyperparameter that is much smaller than the training dataset size, hence the magnitude and the computation time of loss are independent of the number of subjects in the dataset.

Following the completion of the training phase, MGN-Net can map \emph{any} multi-view biological graph to a connectional template that fingerprints the input population. However, each fingerprint is biased towards the particular subject that is used to create it. In order to avoid such bias, we further embed a post-training regularization step (\textbf{Fig.}~\ref{fig:2}\textcolor{blue}{--d}). First, we generate biased population templates by feeding each input to the trained MGN-Net. Then we pick the most centered connections across biased templates by calculating the element-wise median to obtain a finalized connectional template that represents the population most.

% %% ***************************************************************************** %%
\subsection{Multi-view Graph Normalizer Network}
% %% ***************************************************************************** %%
Graphs (i.e. networks) are used extensively in various fields ranging from drug discovery to computational linguistics. They are also the backbone of many biomedical applications due to their great ability to represent knowledge of interacting entities. Despite their ubiquity, graphs cannot be easily used for machine learning applications since there is no straightforward way to encode their non-Euclidean structure into a feature vector representation.  Classical approaches such as handcrafted heuristics \citep{handcrafted}, graph statistics \citep{graph-stats}, and kernel functions \citep{kernels} are widely used to compute feature vectors for predictive tasks on graphs. However, such approaches treat the graph-driven feature vector estimation (i.e. embedding) process as a preprocessing step, therefore, they generally fail to learn task-optimized graph representations that preserve the graph structure and topology. 

The search for data-driven representation learning on graphs ushered in a new deep learning-based approach called graph neural networks (GNNs) \citep{kipf2016, FastLocalized, xu2019powerful, GAttention}. As other conventional methods, GNNs aims to compute a vector embedding for each node in the graph. However, GNNs are unique in the sense that they learn graph embedding through optimizing a predefined function, therefore, they automatically learn the accurate representation that is specific to the problem without any hand engineering. GNNs have recently revolutionized the field of graph theory and network analysis by generalizing  convolutional neural networks \citep{CNNImage} (CNNs), which naturally operate on Euclidean data such as images, to graphs. In particular, CNNs slide small learnable filters through the image and at each position of the filters, small patch of pixels are multiplied by the learnable filters to extract some useful local features of the image such as lines and corners. In deeper layers, these local features are then combined to learn more detailed structural features. By analogy to CNNs working principle, GNNs exploit learnable filters to control how each node aggregates information from its local neighborhood and in each layer, hierarchically merges information passed by the previous layer and neighbors to learn a comprehensive vector representation for each node.

Many GNN based methods can be framed in terms of Message Passing Neural Networks (MPNN) paradigm \citep{MPNN}. Let us represent a graph $G$ with node features $\mathbf{v}_i$ and edge features $\mathbf{e}_{ij}$. The forward pass consists of message passing and the readout phase.  The message passing phase runs for $L$ times (number of graph convolution layers) and is formulated in terms of message functions $M_l$ and node update functions $U_l$. In the message-passing phase, node embeddings $\mathbf{v}_i^{l - 1}$ are updated based on messages $\mathbf{m}_i^{l}$  according to:

\begin{gather}
	\mathbf{m}_i^{l} = \text{AGGR}\left \{ M_l\left ( \mathbf{v}_i^{l-1}, \mathbf{v}_j^{l-1}, \mathbf{e}_{ij} \right ) \right \}_{j \in N(i)}
\end{gather}
\begin{gather}
	\mathbf{v}_i^{l} = U_l\left (\mathbf{v}_i^{l-1}, \mathbf{m}_i^{l}  \right )
\end{gather}

where AGGR is a permutation invariant aggregation function such as mean or max,  $N(i)$ is the set of neighbors of  $i$ in graph G. Next, the readout function $\text{R}$ takes final node embeddings $\mathbf{v}_i^L$, and performs the given task. Finally, a loss computed for the output of $\text{R}$, and the network is trained in an end-to-end fashion.
\begin{gather}
	R\left ( \left \{ \mathbf{v}_i^L \; | \; i \in G  \right \} \right )
\end{gather}
The functions $M_l$, $U_l$, and $R$ are all differentiable functions (at least differentiable everywhere except few points in their domain) so that they can be learned via gradient-based optimization. In what follows, we first formalize our multi-view graph normalization problem then explain the components of our architecture in the frame of MPNN.

We propose the MGN-Net to solve the problem of integrating a population of multi-view networks. This problem can be defined as follows. Let sample $s$ be represented by a set of  $n_v$ weighted undirected graphs with $n_r$ nodes. We model this sample as a single tensor  $\mathcal{T}_{s} \in  \mathbb{R}^{n_{r} \times n_{r} \times n_{v}}$ that is composed of stacked $n_v$ adjacency matrices   $\left \{ \mathbf{X}_s^v  \right \}_{v = 1}^{n_v}$ of $\mathbb{R}^{n_{r} \times n_{r}}$. Our objective is to integrate a set of multi-view graphs $T = \left \{ \mathcal{T}_{1}, \mathcal{T}_{2}, \dots , \mathcal{T}_{N} \right \}$ in order to obtain a  population-representative connectional template $\mathbf{T} \in\mathbb{R}^{n_{r} \times n_{r}}$ that is (1) well-centered, (2) discriminative, and (3) topologically sound. Since there is no general heuristics or conventional methods to meet these three broad constraints in a generic manner, we devised a novel architecture and loss function that learns the best node representation for mapping each sample multi-view graph to a population-representative template using a GNN.  We have summarized the major mathematical notations presented in this paper in \textbf{Table}~\ref{tab:math}.

\begingroup
\begin{table}[H]
	\centering
	\resizebox{\textwidth}{!}{%
	\begin{tabular}{lllllllllllllllll}		
		\hline\noalign{\smallskip}	
		\hline\noalign{\smallskip}
		\multicolumn{3}{l}{Notation} &&&& \multicolumn{11}{l}{Definition}\\
		\hline \hline\noalign{\smallskip}
		
		\multicolumn{3}{l}{$n_r$} &&&& \multicolumn{11}{l}{total number of nodes (regions of interests) in the networks} \\
		\multicolumn{3}{l}{$n_v$} &&&&\multicolumn{11}{l}{total number of measurements (network views) for a sample} \\
		\multicolumn{3}{l}{$\mathcal{T}_s$} &&&& \multicolumn{11}{l}{subject's tensor representation $\in \mathbb{R}^{n_r \times n_r \times n_v}$} \\
		\multicolumn{3}{l}{$T$} &&&& \multicolumn{11}{l}{Training dataset} \\
		\multicolumn{3}{l}{$\mathbf{T}_s$} &&&& \multicolumn{11}{l}{template generated for subject $\mathcal{T}_s$, $\mathbf{T}_s \in \mathbb{R}^{n_r \times n_r}$} \\
		\multicolumn{3}{l}{$\mathbf{T}_{ref}$} &&&& \multicolumn{11}{l}{refined template for training dataset} \\
		\multicolumn{3}{l}{$l$} &&&& \multicolumn{11}{l}{index of a graph convolution layer} \\
		\multicolumn{3}{l}{$\mathbf{e}_{ij}$} &&&& \multicolumn{11}{l}{cross connectional features between node $i$ 
			$j$,  $\mathbf{e}_{ij} \in \mathbb{R}^{n_r \times n_r}$ } \\
		\multicolumn{3}{l}{$F^l$} &&&& \multicolumn{11}{l}{filter-generating network at layer $l$ that maps $\mathbf{e}_{ij}$ to dynamic weights $\mathbf{\Theta}_{ij}^l \in \mathbb{R}^{d_{l} \times d_{l - 1}}$ } \\
		\multicolumn{3}{l}{$\mathbf{v}_i^l$} &&&& \multicolumn{11}{l}{embedding of $i^{th}$ node at layer $l$} \\
		\multicolumn{3}{l}{$N(i)$} &&&& \multicolumn{11}{l}{set of neighbors of node $i$} \\
		\multicolumn{3}{l}{$S$} &&&& \multicolumn{11}{l}{set of indices of randomly selected training samples for loss computation} \\
		\multicolumn{3}{l}{$\mathbf{X}_i^v$} &&&& \multicolumn{11}{l}{$v^{th}$ view of $i^{th}$ random	sample.} \\
		\multicolumn{3}{l}{$\lambda_v$} &&&& \multicolumn{11}{l}{normalization term for view $v$} \\
		\multicolumn{3}{l}{$t_s$} &&&& \multicolumn{11}{l}{normalized node strength distribution of $\mathbf{T}_s$} \\
		\multicolumn{3}{l}{$x_S^v$} &&&& \multicolumn{11}{l}{ground truth normalized node strength distribution for view $v^{th}$ and randomly selected samples $S$ }\\
		\multicolumn{3}{l}{$D_{KL}$} &&&& \multicolumn{11}{l}{Kullback-Liebler divergence} \\
		
		\multicolumn{3}{l}{${\mathcal{L}_{c}}_s^v$} &&&& \multicolumn{11}{l}{centeredness loss for training sample $s$ and view $v$} \\
		\multicolumn{3}{l}{${\mathcal{L}_{kl}}_{v}^s$} &&&& \multicolumn{11}{l}{KL divergence loss for training sample $s$ and view $v$} \\
		\multicolumn{3}{l}{${\mathcal{L}_{tcnl}}_s$} &&&& \multicolumn{11}{l}{overall loss computed for the training sample $s$} \\

		\hline\noalign{\smallskip}
		\hline\noalign{\smallskip}	    
	\end{tabular}}
	
	\caption{\emph{Major mathematical notations used in the paper.} We denote tensors by boldface Euler script letters, e.g., ${\mathcal{X}}$. Matrices are denoted by boldface capital letters, e.g., $\mathbf{X}$ , vectors are denoted by boldface lowercase letters, e.g., $\mathbf{x}$ , scalars and distributions are denoted by lowercase letters, e.g., $x$ ,and sets are denoted by uppercase letters e.g., $X$}
	\label{tab:math}
\end{table}
\endgroup

First we feed sample $s$ represented by a tensor $\mathcal{T}_{s} \in  \mathbb{R}^{n_{r} \times n_{r} \times n_{v}}$  to our MGN-Net architecture (\textbf{Fig.}~\ref{fig:2}\textcolor{blue}{--a}). Then holistic embeddings (i.e. representations) are learned for each node through GNN layers (\textbf{Fig.}~\ref{fig:2}\textcolor{blue}{--b}). These embeddings encapsulate all the complementary information supplied by the different views thanks to the graph convolution operation. There is an abundant variety of graph convolution operations. Particularly for our MGN-Net, we chose edge-conditioned convolution \citep{edgeConditioned},  a graph convolution tailored to simultaneously handle an arbitrary number of different types of weighted edges, which is essential for MGN-Net to blend connectivity patterns across the multiple graph views of each sample.

Given a multi-view graph, let $\mathbf{e}_{ij} \in \mathbb{R}^{n_{v}}$ denote cross-view features between node $i$ and $j$ acquired by stacking edge weights for all the available views such as cortical thickness and sulcal depth. In other words, for each sample $s$, we define the cross-view feature vector  $\mathbf{e}_{ij}$ associated with nodes $i$ and $j$  as  $\left [ \mathcal{T}_s(i,j,1) , \mathcal{T}_s(i,j,2), \dots , \mathcal{T}_s(i,j,n_v) \right ]$. Also, let $l \in \left \{ l_0, \dots, l_{out} \right \}$  be the index of a layer in the architecture and $d_l$  the output dimension of the $l^{th}$ layer. Each layer includes a filter-generating network $F^{l} : \mathbb{R}^{n_v} \mapsto  \mathbb{R}^{d_{l} \times d_{l - 1}}$ that takes $\mathbf{e}_{ij}$ as input and outputs edge-specific weight matrix $\mathbf{\Theta}_{ij}^l$  which dictates the information flow between node $i$ and $j$. This network $F^{l}$ allows the MGN-Net model to learn unique filters at each layer for each pair of nodes (i.e., connection) while exploiting the cross-view edge features.

In terms of the MPNN paradigm, the edge-conditioned convolution defines the  message passing function $M_l$ as:
\begin{gather}
M_l = F^{l}(\mathbf{e}_{ij}; \mathbf{W}^{l}) \mathbf{v}^{l - 1}_{j} + \mathbf{b}^{l} ;  F^{l}(\mathbf{e}_{ij} ; \mathbf{W}^{l}) = \mathbf{\Theta}_{ij}^l
\end{gather}
where $\mathbf{v}{_{i}^{l}} \in \mathbb{R}^{d_{l} \times 1}$  is the embedding of node $i$ in layer $l$ and $\mathbf{b}^{l} \in \mathbb{R}^{d_{l}}$ denotes a network bias and $F^{l}$ is the defined filter-generating network that  maps $\mathbb{R}^{n_{v}}$ to $\mathbb{R}^{d_{l} \times d_{l - 1}}$ with learnable weights $\mathbf{W}^{l}$. Furthermore, node embedding update function $U_l$ defined as:
\begin{gather}
	U_l = \mathbf{\Theta}^{l} .\mathbf{v}{_{i}^{l-1}} + \mathbf{m}_i^{l} ; \mathbf{m}_i^{l} = \text{AGGR}\left \{ M_l\left ( \mathbf{v}_i^{l-1}, \mathbf{v}_j^{l-1}, \mathbf{e}_{ij} \right ) \right \}_{j \in N(i)}
\end{gather}
where $\mathbf{\Theta}^{l} \in \mathbb{R}^{d_{l} \times d_{l - 1}}$ is a learnable parameter, and $N(i)$ denotes the neighbors of node $i$. Noting that our aggregation schema is averaging, the following operation is performed at each layer $l$ for every node $i$.
\begin{gather}
\mathbf{v}{_{i}^{l}} =    \mathbf{\Theta}^{l}  .\mathbf{v}{_{i}^{l-1}} +  \frac{1}{\left |N(i)\right |} \left (\sum_{j \epsilon N(i)} \ F^{l}(\mathbf{e}_{ij}; \mathbf{W}^{l}) \mathbf{v}^{l - 1}_{j} + \mathbf{b}^{l}\right )
\end{gather}

Note that $F^{l}$ can be any type of neural network and vary in each layer depending on the characteristics and complexity of edge weights. Furthermore, $\mathbf{v}_i^{0}$ corresponds to the initial node features of $i$ so this convolution operation can also utilize the node features.  
We note that since brain graphs or connectomes conventionally have no node features, we set entries of the node-specific feature vector $\mathbf{v}_i^{0}$ to `$1$' (i.e., identity vector). Next, we generate an edge-specific weight matrix $\mathbf{\Theta}_{ij}^l$ by learning the filtering function $F^{l}$ for each layer $l$ through optimizing our loss function which will be detailed in the next section. Thanks to filtering functions, each edge from node $i$ to node $j$ is associated with a unique weight matrix (generated by $F^{l}$ from cross-view edge features $\mathbf{e}_{ij}$) which controls the degree of node $i$'s contribution to node $j$'s next layer embedding. Therefore following the first convolution, each node will have a different embedding even though they were identical in the beginning. However, breaking this symmetry of nodes and simultaneously learning distinct node embeddings is a heavy burden to our model. The availability of node features circumvents the need for breaking symmetry, and instead of a noisy input of 1's,  it supplies extra information regarding nodes and their roles in the system. Therefore we foresee that given both node and edge features MGN-Net will output more comprehensive connectional templates and converge to an optimum faster.

After applying three layers of edge conditioned convolution separated by rectified linear unit (ReLU),  we learn abstract embeddings  $\mathbf{V}^{l_{out}} = \left [ \mathbf{v}_1^{l_{out}}, \mathbf{v}_2^{l_{out}}, ..., \mathbf{v}_{n_{r} - 1}^{l_{out}}, \mathbf{v}_{n_{r}}^{l_{out}} \right ]^T$ for the multi-view graph nodes (\textbf{Fig.}~\ref{fig:2}\textcolor{blue}{--b}).  Next, we compute the pairwise relationship of these embeddings to construct the connectional template network (\textbf{Fig.}~\ref{fig:2}\textcolor{blue}{--c}). This step constitutes the readout function $R$ of our model. For our case, we simply used absolute difference since our dataset is composed of morphological dissimilarity networks. The intuition behind this operation is our dataset preparation. We derived our brain networks by computing the pairwise absolute difference in cortical measurements between pairs of ROIs. Since we need to map learned node embeddings (abstract ROIs measurements) back to network representation, we simply repeat our dataset preprocessing step. However, such relationship can be computed by any function that allows for backpropagation so that MGN-Net can be trained in an end-to-end fashion.  For instance, a simple function such as cosine similarity can be used to measure the similarity between output node embeddings. %or artificial neural network can be utilized to learn more complex relationships between node embeddings instead of predefining them. 

% %% ***************************************************************************** %%
\subsection{Loss Function}
% %% ***************************************************************************** %%
Our MGN-Net architecture takes only one sample at a time, however, we aim to output a population-representative connectional template. Since we do not have a ground truth template, to learn the mapping from one sample to the target population-template, we propose to evaluate the output template $\mathbf{T}_s$ that is based on a single sample $s$ against a random subset of the training dataset in the optimization process. In the meantime, we have to preserve the complex topology of the given biological networks while ensuring that the generated template occupies the center of a population by achieving the minimum distance to all population samples (i.e., multi-view networks) to meet our (1) well-centeredness and (3) topological soundness criteria. To address this problem, we present the topology-constrained normalization loss (TCNL) function to evaluate the representativeness of generated templates. TCNL computed against a random subset of training subjects drawn independently for each subject in each epoch. Indices of drawn samples are denoted by the set $S$. The TCNL loss is composed of a weighted sum of two terms. The first term  of the TCNL computes the centredness loss $\mathcal{L}_c$ of the output template and is formalized as follows:

\begin{gather}
{\mathcal{L}_c}_{s}^{v} = \sum_{i \in S} \left \| \mathbf{T}_{s} - \mathbf{X}_{i}^v \right \|_{Frob} \times \lambda_{v}; \; \; \lambda _{v} = \frac{\frac{1}{\mu_{v}}}{\max \left \{   \frac{1}{\mu_{j}} \right \}_{j = 1}^{n_{v}}}
\end{gather}

where $\mathbf{T}_{s}$ denotes the connectional template for the sample $s$ and $\mathbf{X}_i^v$ is the $v^{th}$ view of  $i^{th}$ random sample. $\mu_{v}$ is the mean of  connectivity weights of view $v$ and $\max \left \{  \frac{1}{\mu_j} \right \}_{j = 1}^{n_{v}}$ is the maximum of mean reciprocals $\frac{1}{\mu_{1}}$ to $\frac{1}{\mu_{n_v}}$.   We include an additional view-specific normalization weight $\lambda_{v}$ since the value range of connectional weights for the input graphs might vary largely across views. For example, in our Alzheimer's diseases left hemisphere population (see section \emph{3.1. Evaluation Datasets}), the mean connectional weight for maximum principal curvature is  0.084 with a min-max range in $[0, 0.586]$ while the mean of cortical thickness is 0.723 with a min-max range in $[0, 3.740]$. Therefore, without a normalization term, our MGN-Net model is most likely to overfit the view with the largest connectional weights as it targets to optimize the defined loss function. Similar problems in the literature are addressed by normalizing the adjacency matrices. For instance, SNF \citep{SNF} divides connectivities in each row by the sum of the entries in that row for normalization; however, this breaks the symmetry in the adjacency matrices of the views, therefore, it is not applicable in our case. Moreover, a simple normalization approach such as min-max scaling can saturate connections at 0 and 1 while standard z-score scaling generates negative connectivities in the network that is counter-intuitive for many types of fully positive networks. Therefore, we introduce $\lambda_{v}$ to ensure that the model gives equal attention to each brain view regardless of their value range.

In addition to the centredness loss,  we further add a second loss term $\mathcal{L}_{kl}$ to penalize deviations from the topology of the training networks based on Kullback-Leibler divergence of node strength distribution of the generated connectional brain template from the node strength distribution of randomly selected training samples. Let $\mathbf{A}$ denote the adjacency matrix of a graph. We defined the topological strength $k_i$ of node $i$  and the normalized node strength distribution $n(i)$ as follows:

\begin{gather}
k_i = \sum _j \mathbf{A}_{ij};\; \; n(i) = \frac{k_i}{\sum_{j} k_{j}} 
\end{gather}

First, we compute the normalized node strength distribution of the generated template for sample $s$  which is denoted by  $t_s(i)$. Next, we calculate the ground truth $\left \{  x^1_S(i), ..., x^{n_v}_S(i)\right \}$ for $n_v$ views separately by averaging  normalized node strength distribution of a random subset $S$ of the training population. We calculate the topological loss ${\mathcal{L}_{kl}}_{v}^s$ for subject $s$ and view $v$ as:

\begin{align}
\begin{split}
{\mathcal{L}_{kl}}_{v}^s &= D_{KL}(t_s \left |  \right | x^v_S) +  D_{KL}(x^v_S \left |  \right | t_s )  \\
&= \sum _{i = 1}^{n_r} t_s (i) \log_2\left ( \frac{t_s(i)}{x^v_S(i)} \right )  + \sum _{i = 1}^{n_r} x^v_S(i) \log_2\left ( \frac{x^v_S(i)}{t_s(i)} \right )
\end{split}
\end{align}

Note that the Kullback-Leibler divergence $D_{KL}(P \left |  \right | Q)$ is not a symmetrical function ($D_{KL}(P \left |  \right | Q) \neq D_{KL}(Q \left |  \right | P)$) and defines the information gained by changing beliefs from a prior probability distribution $Q$ to the posterior probability distribution $P$. Intuitively,  $P$ is the true distribution and $Q$ is the estimate. Therefore  $D_{KL}(x^v_S \left |  \right | t_s )$ solely is sufficient to represent the topological loss. However, for our datasets using a symmetrical expression  $D_{KL}(t_s \left |  \right | x^v_S) +  D_{KL}(x^v_S \left |  \right | t_s )$ provides a smoother training.

Given  the training population $T$ and MGN-Net with parameters $\left \{\mathbf{W}_1, \mathbf{b}_1, \dots , \mathbf{W}_{l_{out}}, \mathbf{b}_{l_{out}} \right \}$. We define the TCNL loss $\mathcal{L}_{tcnl_{s}}$ for subject $s$ and the overall optimization task as follows:

\begin{gather}
{\mathcal{L}_{tcnl_{s}}} =   \sum_{v = 1}^{n_v} {\mathcal{L}_c}_s^v + \beta \times {\mathcal{L}_{kl}}_s^v ; \; \min\limits_{\mathbf{W}_1, \mathbf{b}_1 \dots \mathbf{W}_{l_{out}}, \mathbf{b}_{l_{out}}} \frac{1}{\left | T \right |} \sum_{s=1}^{\left | T \right |} {\mathcal{L}_{tcnl_{s}}} 
\end{gather}

% %% ***************************************************************************** %%
\subsection{Post-training Refinement}
% %% ***************************************************************************** %%
MGN-Net learns to map heterogeneous views of each subject to a population-representative template. After MGN-Net training, all learned connectional templates represent the population regardless of the input sample used to generate them. However, each template is biased towards its associated training sample $s$. To eliminate this bias, we suggest an extra step which is executed after the training to obtain more refined and representative templates (\textbf{Fig.}~\ref{fig:2}\textcolor{blue}{--d}). First, each subject is fed through the trained MGN-Net in order to get the corresponding template. Then, the most centered connections are selected from these templates by taking the element-wise median. The median operation could also be replaced with other measures of central tendency such as average or truncated mean, however, we used the centeredness score to empirically verify that the median is the most suitable for our case.

\begin{gather}
\mathbf{T}_{ref} = med \left \{  \mathbf{T}_1, \mathbf{T}_2, \dots, \mathbf{T}_{\left | T \right |} \right \} 
\end{gather}

$\mathbf{T}_{ref}$ denotes the final refined connectional brain template of the input multi-view graph population.
% %% ***************************************************************************** %%
\section{Experiments and Material}
% %% ***************************************************************************** %%

% %% ***************************************************************************** %%
\subsection{Evaluation Datasets}
% %% ***************************************************************************** %%
We showcase MGN-Net with four different evaluation tests: (1) centeredness, (2) topological soundness, (3) accurate identification of most discriminative connections between two biological groups, and (4) biomarker discovery for neurological diseases. We benchmarked our method against SNF \citep{SNF}, netNorm \citep{netnorm} and DGN \citep{DGN} (the ablated version of MGN-Net without the $\mathcal{L}_{kl}$ in the loss function) on a small-scale dataset and a relatively large-scale dataset at the neuroscience scale given that brain disorder datasets are quite scarce. The first datasets (AD/LMCI dataset) consists of 77 subjects (41 subjects diagnosed with Alzheimer's diseases (AD) and 36 with Late Mild Cognitive Impairment (LMCI)) from the Alzheimer's Disease Neuroimaging Initiative (ADNI) database GO public dataset \citep{ADdataset}. Each subject is represented by 4 cortical morphological brain graphs derived from maximum principal curvature, the mean cortical thickness, the mean sulcal depth, and the average curvature measurements \citep{AD-LMCI}. The second dataset (NC/ASD dataset) is collected from the Autism Brain Imaging Data Exchange ABIDE I public dataset dataset \citep{ASDdataset} and includes 310 subjects (155 normal control (NC) and 155 subjects with autism spectral disorder (ASD)) with 6 cortical morphological brain networks extracted from the 4 aforementioned cortical measures in addition to cortical surface area and minimum principle area \citep{soussia2017high,Soussia:2018b}. For each hemisphere, the cortical surface is reconstructed from T1-weighted MRI using the FreeSurfer \citep{freesurfer} pipeline and parcellated into 35 ROIs using Desikan-Killiany \citep{des-kil} atlas and its corresponding brain network is derived by computing the pairwise absolute difference in cortical measurements between pairs of ROIs (\textbf{Table}~\ref{tab:demog}).

\begingroup
\begin{table}[H]
	\centering
	\begin{tabular}{lllllll}		
		\hline\noalign{\smallskip}\hline\noalign{\smallskip}
		Dataset & \multicolumn{2}{c}{AD/LMCI} &  \vline &\multicolumn{2}{c}{NC/ASD} &\\ 
		\cline{2-3} \cline{5-6}
		&  AD   & LMCI      &\vline &  NC   & ASD   \\    \hline
		
		Number of subjects  &41  & 36  &&  155 & 155   \\
		Male  				&23  & 20  &&  124 & 140   \\
		Female 				&18  & 16  &&   31 & 15   \\
		Mean Age  			&75.27  & 72.54  &&  15.36 & 16.92   \\
		Number of Views     &4  & 4   && 6  & 6 \\
		
		\hline\noalign{\smallskip}	
		\hline\noalign{\smallskip}	    
	\end{tabular}
	\caption{\emph{Data distribution for AD-LMCI and NC-ASD datasets.} Each view in the datasets contains 35 nodes, and views are fully connected. In other words, each includes 1190 connections ($35 \times 34$, no self connections). }
	\label{tab:demog}
\end{table}
\endgroup

% %% ***************************************************************************** %%
\subsection{Hyperparameter Setting and Training}
% %% ***************************************************************************** %%
We trained 8 separate models to generate connectional templates for the right and left hemispheres of four groups namely AD, LMCI, NC, and ASD. We set all the hyperparameters for MGN-Net using a grid search. Each model includes 3 edge-conditioned convolution layers while each layer $l$ contains a shallow neural network $F^{l}$ with ReLU activation to map 4 (for AD/LMCI dataset) or 6 (for NC/ASD dataset) connectional features obtained from  heterogeneous views to $\mathbb{R}^{d_{l} \times d_{l - 1}}$ in order to dynamically learn a unique message-passing filter $\mathbf{\Theta}_{ij}^l \in \mathbb{R}^{d_{l} \times d_{l - 1}}$ for each pair of nodes. Furthermore,  edge-conditioned convolution layers output embeddings with 36, 24, and 5 (for AD/LMCI dataset) or 8 (for NC/ASD dataset) dimensions for each node of the input network, respectively. Models are trained using Adam optimizer with a learning rate of 0.0006 and an exponential decay rate of 0.9 for the first moment and 0.99 for the second moment. Since we did not have a GPU memory bottleneck, we executed an update once for the epoch utilizing the gradients computed for the whole dataset (batch size = dataset size). We set the size of the random subset of training samples in our TCNL function to 10 and $\beta$ which balances the $\mathcal{L}_c$\ and $\mathcal{L}_{kl}$  to 25 for AD-LMCI and 10 for NC-ASD. The algorithms are implemented using  PyTorch and PyTorch-Geometric \citep{Fey/Lenssen/2019} frameworks. 

We split each group into training and testing sets using 5-fold cross-validation which yields 80\% to 20\% split. We let each model train for a maximum of 1200 epoch and apply early stopping if there is no improvement in the testing loss for more than 50 epochs.  We note that during the training there was neither a significant difference between training and testing losses nor a sign of overfitting due to excessive training. This can be explained by our randomization process during loss calculation which perturbs the training distribution by selecting a different subset for each optimization step. For instance, given our smallest group  LMCI (37 subjects), there are more than $\binom{37}{10}  \approx 3.4 \times 10^{8}$ different targets for MGN-Net to fit which makes overfitting much harder compared to conventional learning tasks where targets are fixed.

% %% ***************************************************************************** %%
\subsection{Evaluating the Topological Soundness of the Learned Connectional Templates}
% %% ***************************************************************************** %%

Based on convergent evidence from empirical studies that reveal topological patterns in complex networks \citep{CBN,transportation,smallworld}, we suggest that the topology of networks should be preserved during the integration process to obtain more holistic and representative templates. There are many studies investigating the topological features of structural and functional brain networks such as small-world topology, highly connected hubs, and modularity at both the whole-brain scale of human neuroimaging and at a cellular scale in non-human animals \citep{CBN}. Here we assume that imposing the simple node strength distribution constraint on the MGN-Net training is sufficient to preserve the population topology to a large extent since more complex topological measures such as PageRank and effective size are derived from node strength.

As for the evaluation of the topological soundness of the learned brain connectional templates, we used additional topological measures including PageRank \citep{pagerank}, effective size \citep{effectiveSizel}, and weighted clustering coefficient \citep{clustering} measures as they capture different topological properties of graphs. For each measure, we computed the Kullback-Liebler divergence of generated template distribution from the ground truth distribution ($D_{KL}(g \left |  \right | t )$ where $t$ denotes the template measure distribution and $g$ is the ground truth measure distribution). We applied 5-fold cross-validation and generated connectional templates using the \emph{training samples} using MGN-Net, SNF, netNorm, and DGN, respectively.  Next, we computed the aforementioned measures on the connectional templates generated by all three methods. We acquired ground truth by averaging measures that are independently calculated for each view of each \emph{testing sample}. Finally, we reported the average Kullback-Liebler divergence across folds between the generated template distribution and the ground truth distribution. Note that we normalized each distribution using the sum of all node measures to get a valid discrete probability distribution (see \emph{4.3.Topological Soundness Test} for benchmark results) .

\subsubsection{PageRank}
Originally PageRank algorithm is proposed to measure the importance of website pages based on a graph that is derived from World Wide Web pages and hyperlinks between them. However, it is a general measure which can be applied to investigate graph topologies in various domains including biology and neuroscience \citep{BeyondWeb}. We used the power iteration method for calculation and set the damping parameter and maximum iteration to 0.85 and 100, respectively.

\subsubsection{Effective Node Size}
A node's ego network consists of its direct neighbors plus the ties among these neighbors. The effective size of a node measures the non-redundancy of the node's ego network \citep{effectiveSizel}. This measure is formulated as:

\begin{gather}
e(i) = \sum_{j \in N\left ( i \right ) - \left \{ i \right \}} \left ( 1 - \sum _{k \in N(j)} p_{ik} m_{jk} \right )
\end{gather}

where $N(i)$ is set of neighbors of $i$, $p_{ik}$ is the proportion of the weight of the edge connecting node $i$ to node $j$  to the sum of all connection weights of node $i$. $m_{jk}$ denotes the $j$'s connection with $k$'s divided by the $j$'s strongest connection.

\subsubsection{Clustering Coefficient}
There are multiple ways to define the clustering coefficient for weighted graphs. The one we used for our experiments utilizes the geometric average of edge weights of the triangles through the subject node \citep{clustering}: 

\begin{gather}
c(i) = \frac{1}{deg(i) \left ( deg(i) - 1 \right )} \sum_{j, k \in Tri(i)}\left ( \hat{w_{ij}}  \hat{w_{ik}} \hat{w_{jk}} \right )^{1/3}
\end{gather}

$deg(i)$ is the degree of node $i$, $Tri(i)$ is the set of node pairs that form a triangle with node $i$ and  $\hat{w_{ij}}$ is the weight of the connection between node $i$ and $j$ normalized by the maximum connectivity weight in the graph.

\begin{figure}[H]
	\centering
	\includegraphics[width=14cm]{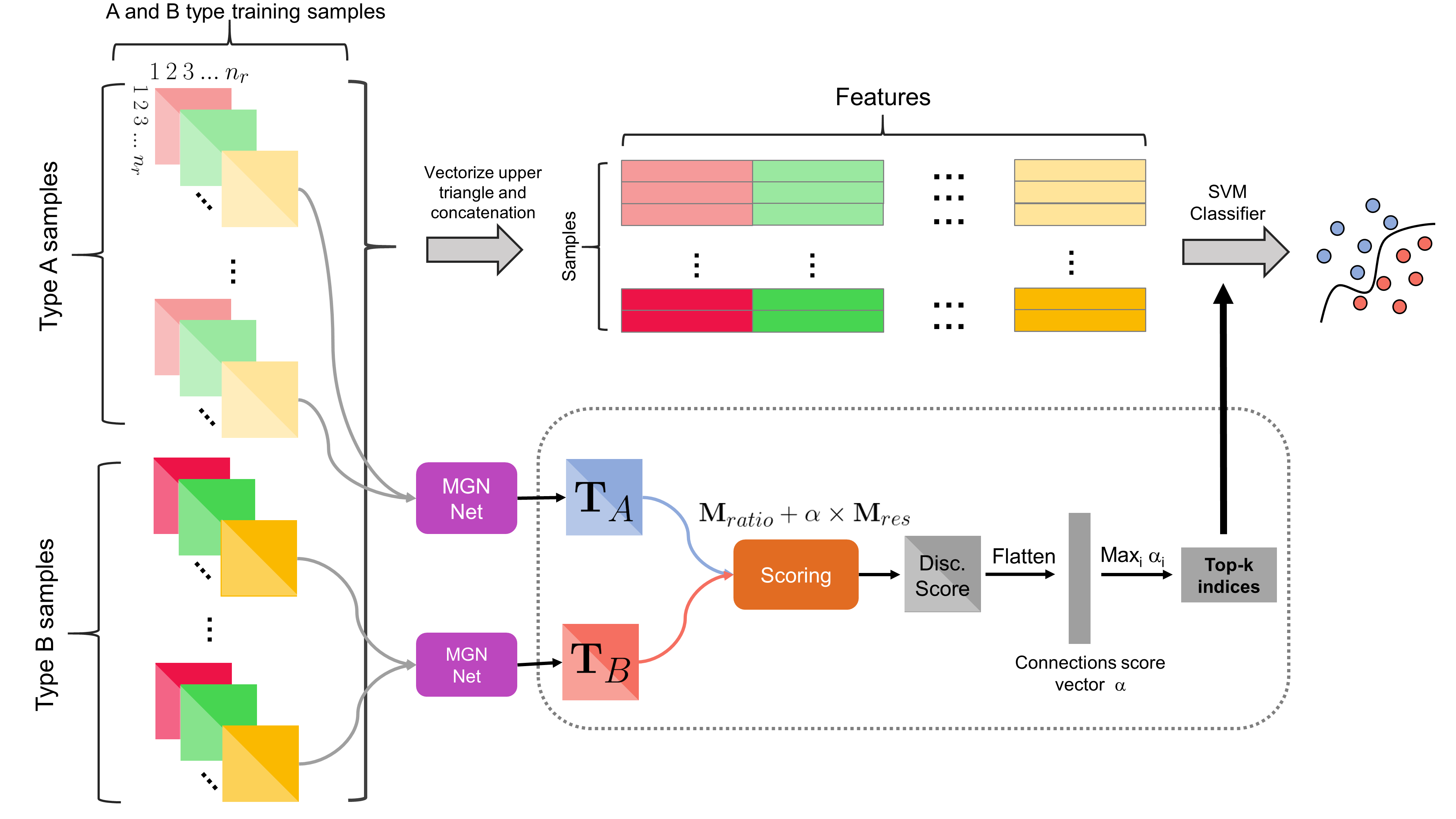}
	\caption{\emph{Discriminative feature selection pipeline for classifying brain networks using templates.} First, we generate two distinct connectional templates for type $A$ samples and type $B$ samples. Then we assign a  discriminativeness score for each connection using both residual matrix $\left \| \mathbf{T}_{A} - \mathbf{T}_{B} \right \|_{Frob}$ and absolute alteration ratio matrix $\max \left ( \frac{\mathbf{T}_{A}}{\mathbf{T}_{B}}, \frac{\mathbf{T}_{B}}{\mathbf{T}_{A}} \right )$. Next, we pick the top $k$ connections to train a support vector machine (SVM) model for classifying new testing samples as $A$ or $B$. 
	}
	\label{fig:disc}
\end{figure}

% %% ***************************************************************************** %%
\subsection{Discriminative Feature Selection and Biomarker Discovery}
% %% ***************************************************************************** %%
Our second criterion is that the generated templates are discriminative which means that the templates will encapsulate the most distinctive traits of the input population. Note that our loss function does not include any term to enforce this criterion and that our model training is performed on a single population. However, through our feature selection and biomarker discovery experiments, we showed that the MGN-Net integration process fulfills this criterion. %First, we utilized the learned templates by MGN-Net, SNF, and netNorm to separately pick the most discriminative features (i.e., brain connections) for classifying disordered and healthy brain states. Next, we showed that SVM with feature selection guided by MGN-Net's connectional templates outperforms others. Second, we demonstrated that there is a prominent correlation between discriminative features pinpointed by MGN-Net connectional templates and previous works that study biomarkers for AD, LMCI, and ASD.

Specifically, we propose to compare templates ($\mathbf{T}_{A}$ and $\mathbf{T}_{B}$) generated from two groups $A$ and $B$ to identify the most discriminative connections disentangling both groups. To do so, we calculate a discriminativeness score for each graph connection by taking both residual matrix $\left \| \mathbf{T}_{A} - \mathbf{T}_{B} \right \|_{Frob}$ and absolute alteration ratio matrix $\max \left ( \frac{\mathbf{T}_{A}}{\mathbf{T}_{B}}, \frac{\mathbf{T}_{B}}{\mathbf{T}_{A}} \right )$ into consideration. Alteration ratio plays a key role in giving equal attention to connections with relatively small weights since the residual matrix alone focuses on the deviation of larger connections. Next, we select the edges (brain connections) with the highest discriminativeness scores:

\begin{gather}
Score_{ij} = \max \left ( \frac{\mathbf{T}_{A_{ij}}}{\mathbf{T}_{B_{ij}}}, \frac{\mathbf{T}_{B_{ij}}}{\mathbf{T}_{A_{ij}}} \right ) + \alpha \times \left \| \mathbf{T}_{A_{ij}} - \mathbf{T}_{B_{ij}} \right \|_{Frob}
\end{gather}

Here $\alpha$ is a parameter to balance residual and alteration values for computing the discriminativeness score. We define it as $\frac{2}{\mu_{\mathbf{T}_{A}} + \mu_{\mathbf{T}_{B}}}$  where $\mu_{\mathbf{T}_{A}}$ is the average of all connection weights of connectional template $\mathbf{T}_{A}$. Note that division by zero may occur during the computation of $\max \left ( \frac{\mathbf{T}_{A}}{\mathbf{T}_{B}}, \frac{\mathbf{T}_{B}}{\mathbf{T}_{A}} \right )$; we simply set alteration ratio to zero for such cases.

For classifying populations (\textbf{Fig.}~\ref{fig:disc}), we first pick the top $k$ connections with the highest discriminativeness scores calculated using two connectional templates generated from the training samples. Next, we concatenate the cross-view edge features $\mathbf{e}_{ij}$ of the selected connections, so we obtain $k \times 4$ features for AD-LMCI  populations and $k \times 6$ features for NC-ASD populations. Finally, we train a support vector machine (SVM) classifier on these features to perform binary classification. We present the average accuracy across both 5-folds and $k$ values where $k = \left \{5, 10, \dots, 25  \right \}$. Note that we perform 16 classifications in total  (4 populations namely,  AD-LMCI left hemisphere, AD-LMCI right hemisphere, NC-ASD left hemisphere, and NC-ASD right hemisphere repeated for MGN-Net and three baseline methods) and we independently select the optimal hyperparameters such as kernel type, kernel-specific hyperparameters, regularization parameter $C$, and tolerance for stopping criterion for each classification task using grid search (over $13600$ combinations) by choosing the hyperparameters that yield best average accuracy on the 5-fold cross-validation. SVM classification, cross-validation, and grid search procedures were implemented using Scikit-learn \citep{scikit-learn}.

% %% ***************************************************************************** %%
\subsection{Remarks on the SNF, netNorm and DGN Benchmarks}
% %% ***************************************************************************** %%
Originally SNF \citep{SNF} is applied to fuse similarity networks of cancer patients for clustering applications for cancer subtyping.  However, it is mathematically demonstrated that SNF can also be used to fuse dissimilarity networks where connections denote the difference between nodes as in our datasets \citep{netnorm}. Furthermore, SNF is tailored to integrate a population of \emph{single-view} networks whereas our multi-view network integration task operates at two levels: multi-view network integration of individuals and population integration across subjects. Therefore, we adapt SNF to our problem by combining it with averaging. We tried three different pipelines SNF-SNF (SS), Average-SNF (AS), and SNF-Average (SA) where the first step is the merging of the multi-view network for each individual, and the second one is the fusion across subjects. For clarity,  we only present the results by SA as it largely outperformed the 3 other SNF-based alternatives across all evaluation datasets. As for netNorm \citep{netnorm} and the DGN (ablated version) \citep{DGN},  we rely on the publicly available implementations\footnote{\url{https://github.com/basiralab/netNorm} \& \url{https://github.com/basiralab/DGN}} and no adjustments are made since they are directly applicable to our datasets.

% %% ***************************************************************************** %%
\section{Results}
% %% ***************************************************************************** %%

% %% ***************************************************************************** %%
\subsection{Centeredness Test}
\begin{figure}[H]
	\centering
	\includegraphics[width=14cm]{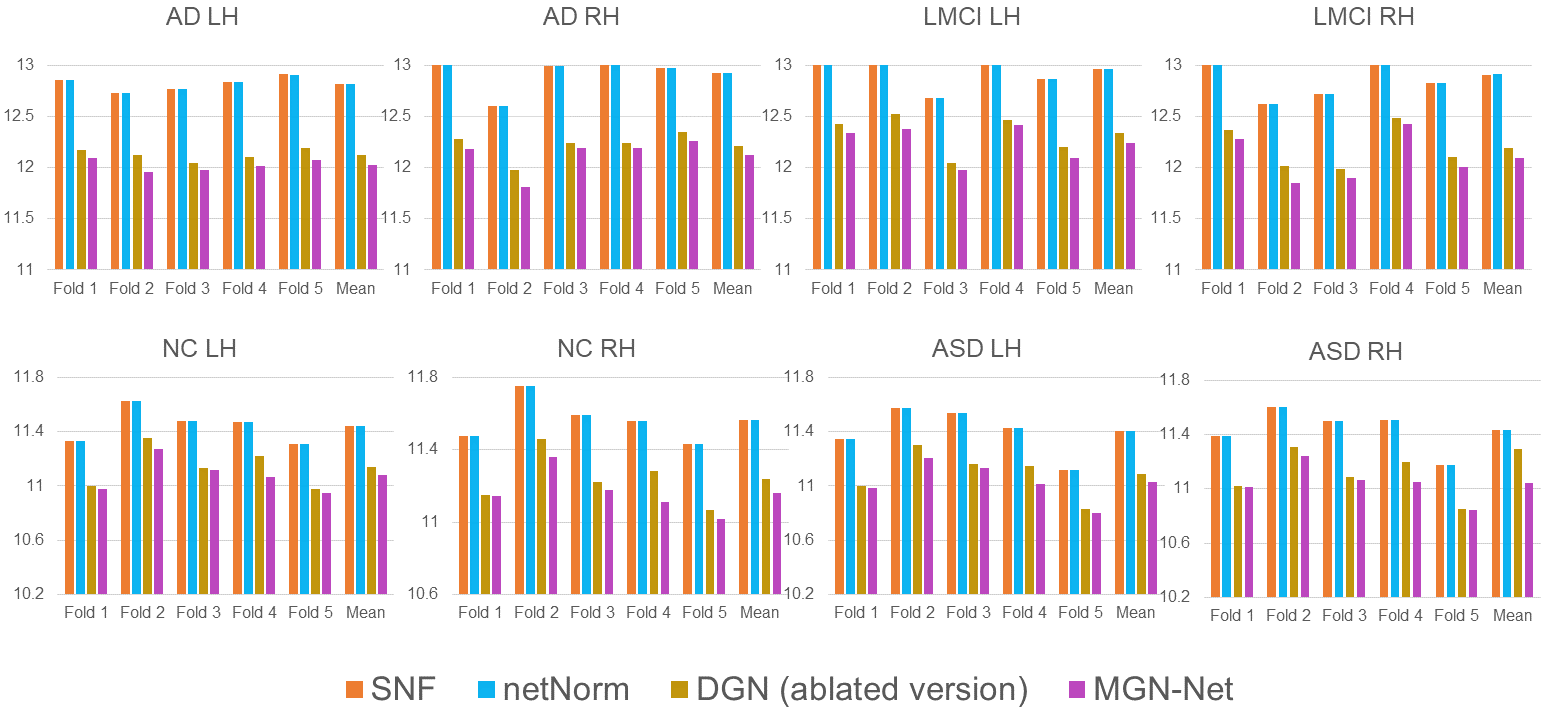}
	\caption{\emph{Centeredness comparison of CBTs generated by SNF \citep{SNF}, netNorm \citep{netnorm}, DGN (ablated version) \citep{DGN} and MGN-Net.}  Charts illustrate the mean Frobenius distance between the connectional templates created using the training sets and networks of the samples in the testing set using a 5-fold cross-validation strategy. We report the average distance for each cross-validation fold as well as the average across folds (``Mean'' bars on the right). MGN-Net achieved the lowest mean Frobenius distance to the population multi-view networks with a high statistical significance demonstrated by a two-tailed paired t-test (all $p > 0.0001$) for MGN-Net-SNF and MGN-Net-netNorm pairs. As for the MGN-Net-DGN pair, all $p > 0.05$ except NC LH and ASD RH. LH: left hemisphere. RH: right hemisphere.}
	\label{fig:3}
\end{figure}

% %% ***************************************************************************** %%
For reproducibility and generalizability, we split the datasets into training and testing sets using 5-fold cross-validation. We used the training set to generate the CBTs for the population and compute the centeredness on the left out testing fold. To evaluate the centeredness of the integrated networks (CBTs), we measure the mean Frobenius distance between each network view of each sample in the testing left-out fold and the CBT of the four training folds. We note that MGN-Net significantly outperforms other methods across all folds and evaluation datasets (\textbf{Fig.}~\ref{fig:3}, two-tailed paired t-test, $p > 0.05$, see \textbf{Table}~\ref{tab:stdcentredness} for standard deviation across folds).

\begingroup
\begin{table}[H]
	\centering
	\begin{tabular} {llllll}
		\hline\noalign{\smallskip}\hline\noalign{\smallskip}
		Dataset & \multicolumn{4}{c}{Centeredness standard dev. across folds}  \\  \hline
		&  SNF   & netNorm  & DGN &  MGN-Net \\    \hline
		AD LH  & 0.07 &  0.07 & 0.05 & 0.06\\
		AD RH  & 0.19 &  0.18 & 0.14 & 0.18\\
		
		LMCI LH  & 0.19 &  0.19 & 0.2 & 0.19\\
		LMCI RH  & 0.28 &  0.27 & 0.22 & 0.25\\
		
		NC LH  & 0.13 &  0.13 & 0.15 & 0.13\\
		NC RH  & 0.12 &  0.12 & 0.15 & 0.12\\
		
		ASD LH  & 0.18 &  0.18 & 0.18 & 0.15\\
		ASD RH  & 0.16  &  0.16 & 0.18 & 0.14\\
		\hline\noalign{\smallskip}	
		\hline\noalign{\smallskip}
	\end{tabular}
	\caption{\emph{Standard deviations of Frobenius distances between the connectional brain templates created using the training sets and networks of the samples in the testing set across 5 cross-validation folds.}}
	\label{tab:stdcentredness}
\end{table}
\endgroup

\begin{figure}[H]
	\centering
	\includegraphics[width=11cm]{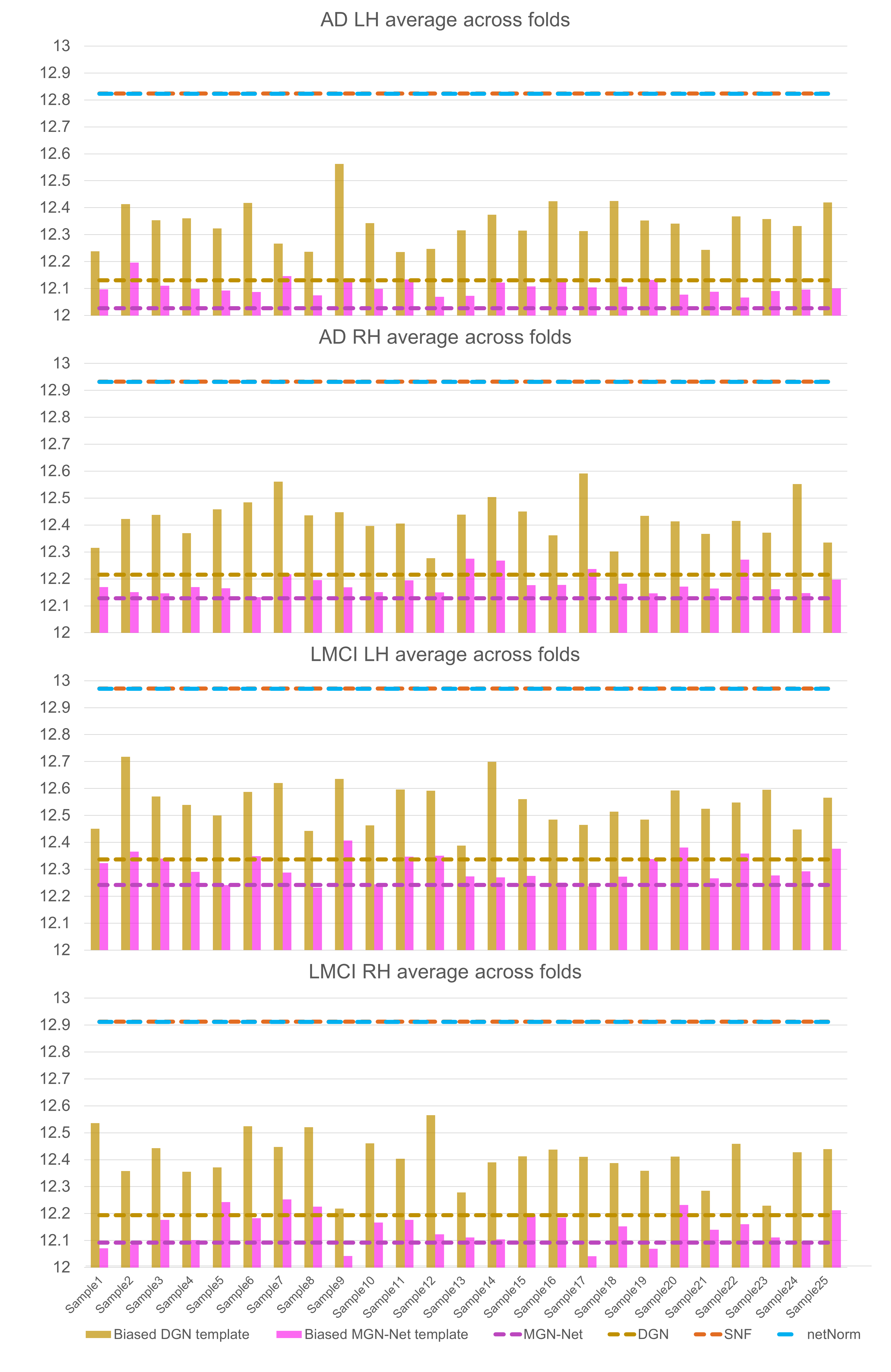}
	\caption{This chart displays the average centeredness distance across folds of the learned AD and LMCI templates that are biased towards randomly selected 25 samples along with the finalized median-based templates generated by MGN-Net, DGN (ablated version) \citep{DGN}, SNF \citep{SNF}, and netNorm \citep{netnorm}. AD: Alzheimer's disease. LMCI: late mild cognitive impairment.
	}
	\label{fig:s12}
\end{figure}

\begin{figure}[H]
	\centering
	\includegraphics[width=11cm]{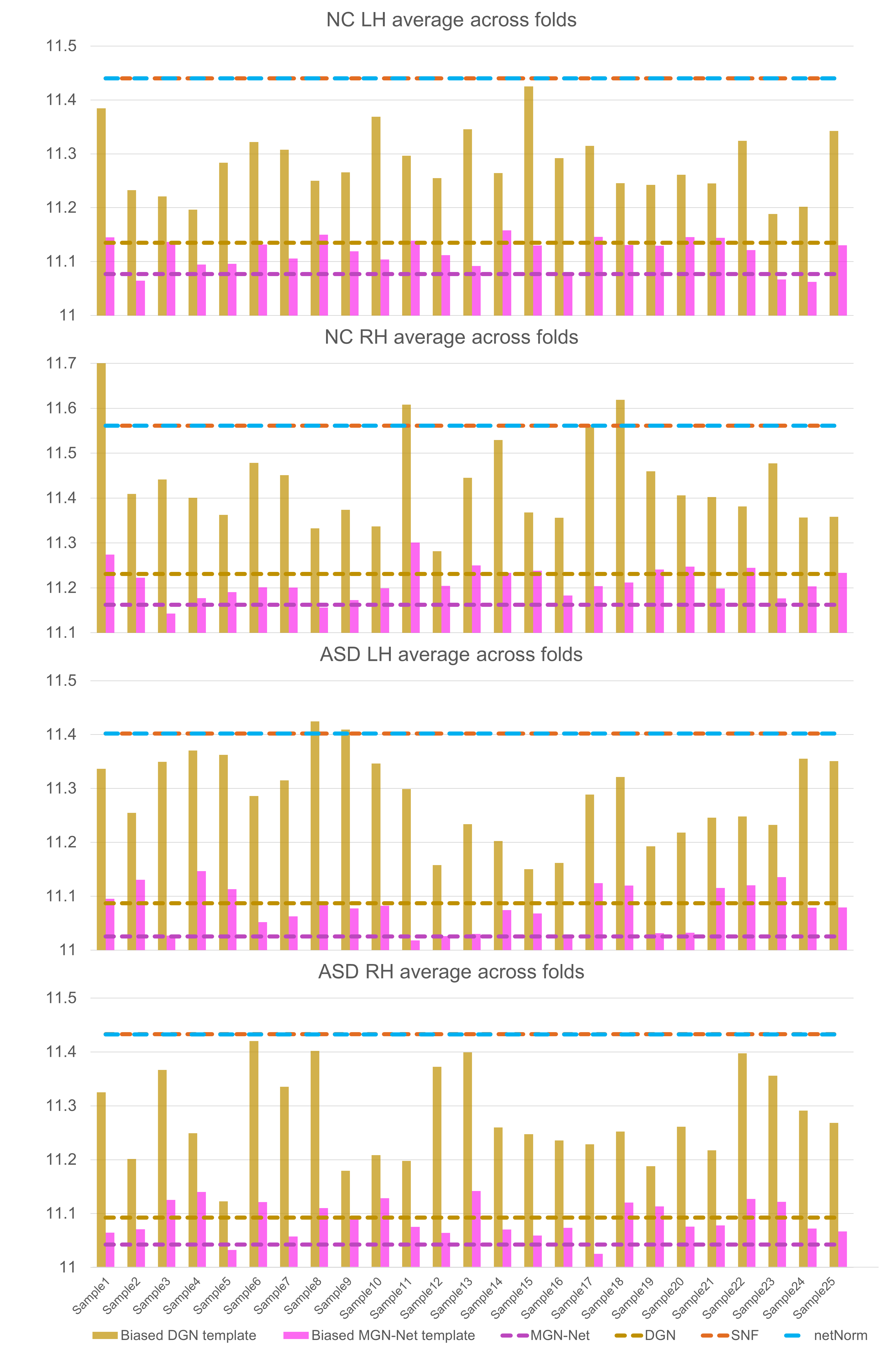}
	\caption{This chart displays the average centeredness distance across folds of the learned NC and ASD templates that are biased towards randomly selected 25 samples along with the finalized templates generated by MGN-Net, DGN (ablated version)  \citep{DGN}, SNF \citep{SNF}, and netNorm \citep{netnorm}. NC: normal controls. ASD: autistic spectrum disorder.
	}
	\label{fig:s13}
\end{figure}

% %% ***************************************************************************** %%
\subsection{Centeredness of Subject Biased Templates}
% %% ***************************************************************************** %%
 We run validation experiments only on the refined connectional templates produced by taking the element-wise median of subject-biased templates to further refine the learned templates by pruning biased connections. Here, we take a look at the centeredness of randomly selected $25$ subject-biased CBTs. We note that even the subject-biased templates (i.e. preliminary outputs) learned by the MGN-Net significantly outperform templates that are produced by SNF and netNorm in terms of centeredness (\textbf{Fig.}~\ref{fig:s12}, \textbf{Fig.}~\ref{fig:s13}).

Although it was not the motivation behind the topology-constrained loss function, penalizing deviations from the node strength distributions of the population provided much more consistent subject biased templates. For example, the standard deviation of the centeredness score for randomly selected subject-biased templates (AD LH. population) estimated by DGN is $0.077$ whereas for the MGN-Net it reaches only $0.029$. This phenomenon can be explained by the fact that preserving the population topological properties regularizes the subject-biased templates by avoiding motifs and connections that are not repeated across the population.

\begin{figure}[H]
	\centering
	\includegraphics[width=14cm]{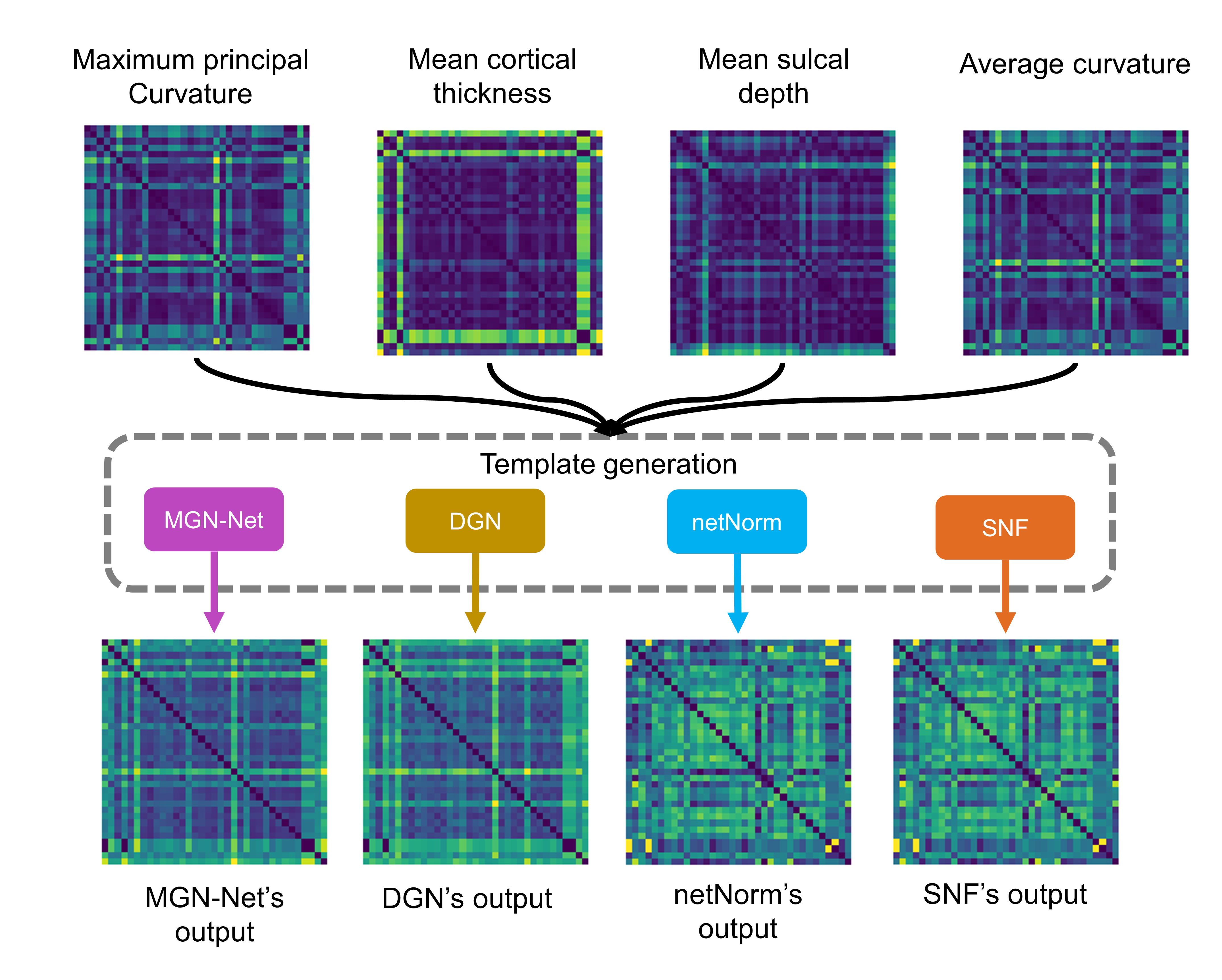}
	\caption{\emph{The figure demonstrates templates estimated by MGN-Net, DGN (ablated version) \citep{DGN}, netNorm \citep{netnorm}, and SNF \citep{SNF} for the left hemisphere of the AD group along with view averages.}  It is apparent that the templates generated by MGN-Net and DGN encapsulates topological patterns which commonly exist in all views.  As for netNorm and SNF case, they capture only a few local motifs across views. }
	\label{fig:s8}
\end{figure}

% %% ***************************************************************************** %%
\subsection{Topological Soundness Test}
% %% ***************************************************************************** %%
 Network science provides us with tools to quantify the topological properties of networks such as clustering, structural holes, and centrality. It is shown that distinctive topological patterns occur in a wide range of complex networks from biological to social networks \citep{CBN,smallworld,transportation}. Therefore, it is essential to preserve these complex but systematic patterns during the biological network data integration process to generate more realistic and integral templates. Thanks to the proposed topology-constrained loss function (TCNL), we can preserve any topological pattern in a population of multi-view networks when transforming them to a unique holistic connectional template --in a fully generic manner (\textbf{Fig.}~\ref{fig:s8}).
 
\begin{figure}[H]
	\centering
	\includegraphics[width=13cm]{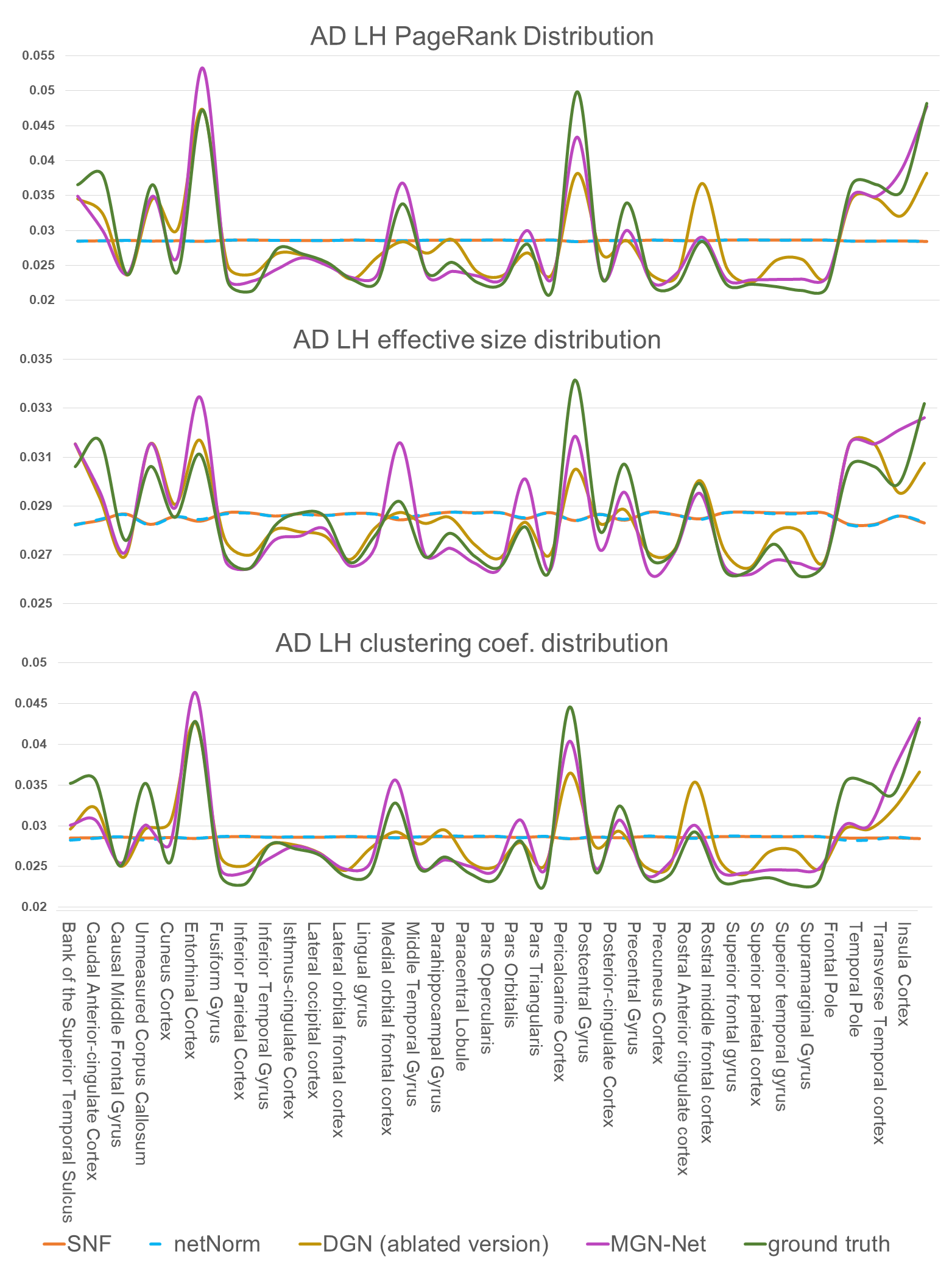}
	\caption{\emph{Comparison of topological distributions of templates generated by SNF \citep{SNF}, netNorm \citep{netnorm}, DGN (ablated version) \citep{DGN}, and MGN-Net against the ground truth network population distribution for the AD left hemisphere population.} AD: Alzheimer's disease.}
	\label{fig:top_test}
\end{figure}

 In the context of network neuroscience, we evaluated the topological soundness of the learned CBTs by comparing the discrepancy of the distribution of various topological measures including PageRank \citep{pagerank}, effective node size \citep{effectiveSizel}, and clustering coefficient \citep{clustering} between the population multi-view brain networks and the estimated CBT. First, we calculated the ground truth by averaging the distribution of topological measures (e.g., clustering coefficient) of each network view of each testing subject. Next, we computed the distribution of topological measures for the CBT estimated using the training dataset. Specifically, each distribution is a discrete probability distribution that is composed of topological measures calculated for each node normalized by the total sum of measures across all nodes. Lastly, we computed the Kullback-Leibler divergence of the ground truth distributions and distributions derived from the connectional brain templates. These steps were completed with a comprehensive battery of graph topology analysis, to assess the consistency and generalizability of the new TCNL function where we demonstrated that a simple node strength constraint is sufficient for endowing MGN-Net with the ability to capture much more complex topological measures such as PageRank and effective node size. We display the PageRank, effective size and clustering coefficient results in the form of distribution graphs for AD LH population (\textbf{Fig.}~\ref{fig:top_test}) and also in a tabular form for all populations (\textbf{Table}~\ref{tab:1}). Remarkably, the connectional brain template generated by our MGN-Net architecture captures the topology of the connectomic datasets by showing striking similarity with the ground truth while netNorm and SNF fail to preserve the multi-view connectomic data topology. Our statistical analysis using two-tailed paired t-test also demonstrates that MGN-Net significantly outperformed other methods on both small-scale and large-scale evaluation datasets and across all topological measures (two-tailed paired t-test $p < 0.005$).

\begingroup
\begin{table}[H]
	\centering
	{\renewcommand{\arraystretch}{1.2} 
		\resizebox{\textwidth}{!}{%
			\begin{tabular}{lllllllllllllll}
				\hline\noalign{\smallskip}\hline\noalign{\smallskip}
				
				Dataset & \multicolumn{4}{c}{PageRank Dis. \textcolor{red}{*}} &  \vline &\multicolumn{4}{c}{Effective Size Dis. \textcolor{red}{*}} &  \vline & \multicolumn{4}{c}{Clustering Coef. \textcolor{red}{*}}  \\ 
				\cline{0-14}
				&  SNF   & netNorm  &  DGN & \bf{MGN-Net}   &\vline &  SNF     & netNorm &  DGN  & \bf{MGN-Net}   &\vline  &  SNF     & netNorm &  DGN   & \bf{MGN-Net} \\    \hline
				
				AD LH  &0.0552  &0.0552 & 0.0123 & \bf{0.0046} &&  0.0042 & 0.0041 & 0.0010 & \bf{0.0009} && 0.0333 & 0.0333 & 0.0099  &\bf{0.0049}  \\
				
				LMCI LH  &0.0560  &0.0560 & 0.0147 & \bf{0.0071} &&  0.0041 & 0.0040 & 0.0011 & \bf{0.0010} && 0.0338 & 0.0338 & 0.0109 &\bf{0.0060}  \\
				
				AD RH  &0.0578  &0.0578 & 0.0121 & \bf{0.0007} &&  0.0038 & 0.0037 & 0.0008 & \bf{0.0005} && 0.0348 & 0.0348 & 0.0099  &\bf{0.0029}  \\
				
				LMCI RH  &0.0588  &0.0588 & 0.0117  &\bf{0.0007} &&  0.0038 & 0.0037 & 0.0008 &\bf{0.0005} && 0.0354 & 0.0354 & 0.0097 & \bf{0.0029}  \\
				
				NC LH  &0.0599  &0.0600 &  0.0104& \bf{0.0039} &&  0.0104 & 0.0104 &0.0015 &\bf{0.0012} && 0.0341 & 0.0341 & 0.0114 & \bf{0.0055}  \\
				
				ASD LH  &0.0573  &0.0573 &  0.0100 & \bf{0.0040} &&  0.0100 & 0.0100 & 0.0012  &\bf{0.0012} && 0.0326 & 0.0327 & 0.107 & \bf{0.0056}  \\
				
				NC RH  &0.0629  &0.0630 &  0.0098& \bf{0.0011} &&  0.0102 & 0.0102 & 0.0013 &\bf{0.0008} && 0.0361 & 0.0362 & 0.0110 & \bf{0.0039}  \\
				
				ASD RH  &0.0625  &0.0625 &  0.0087& \bf{0.0013} &&  0.0101 & 0.0101 & 0.0010 &\bf{0.0007} && 0.0359 & 0.0359 & 0.0102 &\bf{0.0043}  \\
				\hline\noalign{\smallskip}	
				\hline\noalign{\smallskip}	    
	\end{tabular}}}
	\caption{\emph{Evaluation of deviation from ground truth topology.} We report the Kullback-Liebler divergence of the ground truth and learned connectional templates for the PageRank, effective size, and clustering coefficient distributions. \textcolor{red}{*} $p < 0.005$ for MGN-Net \emph{vs} SNF, MGN-Net \emph{vs} netNorm and MGN-Net \emph{vs} DGN (ablated version) for all metrics using two-tailed paired t-test.}
	\label{tab:1}
\end{table}
\endgroup

% %% ***************************************************************************** %%
\subsection{Discriminative Feature Selection Test}
% %% ***************************************************************************** %%
Having demonstrated that MGN-Net generates both well-centered and topologically sound connectional templates, we next evaluated its capacity to preserve distinctive features (i.e., edge/connection weights) through the normalization process. We reason that a well-representative connectional template can encapsulate the most unique traits of a population of multi-view networks, which makes it easily distinguishable from other population templates. Those uniquely distinctive features can be used for biological network classification tasks such as differentiating the healthy from the disordered brain network. We designed a simple strategy based on the learned templates that automatically identify the most discriminative network connections distinguishing between two populations. First, we used integration methods to generate two CBTs for populations $A$ and $B$, respectively. Then we assigned a discriminativeness score to each brain connectivity between a pair of ROIs based on the high peaks in the difference (weighted sum of the alteration ratio and magnitude) between the adjacency matrices of connectional templates $A$ and $B$. The intuition behind this comparison is driven by our hypothesis that samples belonging to a population $A$ lie further away from the CBT of population $B$ in comparison with their induced CBT (i.e., from population $A$) and vice versa. Hence, a simple comparison between the connectivity matrices of the CBTs of populations $A$ and $B$ can easily reveal the most distinctive connectivities to be used in boosting the accuracy of an independent classification method (\textbf{Fig.}~\ref{fig:disc}).

\begingroup
\begin{table}[H]
	\centering
	\begin{tabular} {llllll}
		\hline\noalign{\smallskip}\hline\noalign{\smallskip}
		Dataset & \multicolumn{4}{c}{Ave. Accuracy}  \\  \hline
					&  SNF   & netNorm  & DGN &  \bf{MGN-Net} \\    \hline
		AD-LMCI LH  &70.92 ($\sigma \; 1.63$)  & 71.37  ($\sigma \; 0.93$) & 64.63 ($\sigma \; 3.30$) &\bf{74.23} ($\sigma \; 2.91$) \\
		AD-LMCI RH  &\bf{56.72}  ($\sigma \; 0.73$) & 55.92 ($\sigma \; 0.93$) & 53.25 ($\sigma \; 0.00$) &  54.05 ($\sigma \; 1.19$) \\
		NC-ASD LH  &54.58  ($\sigma \; 1.06$) & 54.97 ($\sigma \; 1.87$) & 55.94 ($\sigma \; 0.54$) & \bf{56.26} ($\sigma \; 0.58$) \\
		NC-ASD RH  &55.45  ($\sigma \; 0.77$) & 56  ($\sigma \; 1.56$)& 57.16 ($\sigma \; 1.52$) & \bf{58.13} ($\sigma \; 1.05$) \\
		\hline\noalign{\smallskip}	
		\hline\noalign{\smallskip}
	\end{tabular}
	\caption{\emph{SVM binary classification results using features selected by  SNF \citep{SNF}, netNorm \citep{netnorm}, DGN (ablated version) \citep{DGN} and MGN-Net.} Average accuracy across 5 folds and $k$ values are presented for classifying AD-LMCI and NC-ASD using the left and right hemispheres features separately. The standard deviations $\sigma$ of average accuracies across k values are indicated in parenthesis.}
	\label{tab:2}
\end{table}
\endgroup

As a proof of concept for the validity of our hypothesis, we further explored the discriminative power of the learned CBTs by our MGN-Net architecture in comparison with benchmark methods. We demonstrated the reproducibility of our classification results against different perturbation of the training and testing sets using 5-fold cross-validation and across both AD and ASD neurological disorders. 
We set up 4 different classification tasks namely; AD-LMCI left hemisphere, AD-LMCI right hemisphere, NC-ASD left hemisphere, and NC-ASD right hemisphere. Next, we selected the top $k$ ($k = \left \{5,10, \dots, 25 \right \}$) most discriminative connections revealed by the learned connectional templates of each method and fed them to a support vector machines (SVM) classifier for training. Note that by setting small $k$ values, we are only using less than one percent of the available features. Also, note that brain network classification is a very challenging task that requires deliberately designed preprocessing steps and architectures to achieve satisfactory accuracy. Here, our very modest pipeline that consists of template-based feature selection and SVM is designed to evaluate the templates' capability of identifying important connections. We empirically set the best SVM parameters including the kernels for each method independently using grid search. \textbf{Table}~\ref{tab:2} displays the average accuracy across folds and selected $k$ values. Remarkably, the SVM with training features selected by our MGN-Net outperformed baseline methods on 3 classification tasks --except the AD-LMCI (RH). These results imply that our proposed model not only generates more centered and topologically sound connectional templates but is significantly better at capturing unique traits of multi-view graph populations. MGN-Net outperformance was also replicable in different multi-view connectomic datasets with subtle connectional changes between comparison groups, suggesting that our model could successfully spot integral and holistic brain connections that largely vary across both populations --which leads us to the next series of experiments on population template fingerprinting.

\begin{table}[H]
	\resizebox{\textwidth}{!}{%
		\begin{tabular}{c|c|c}
			\hline \hline
			Rank & AD-LMCI LH                                                              & AD-LMCI RH                                                            \\ \hline\hline
			1    & \textbf{Entorhinal Cortex}   $\longleftrightarrow$   Superior Temporal Sulcus     & \textbf{Cuneus Cortex}   $\longleftrightarrow$ Fusiform Gyrus
			\\ \hline
			2    & \textbf{Entorhinal Cortex}     $\longleftrightarrow$   Unmeasured Corpus Callosum     & \textbf{Cuneus Cortex}   $\longleftrightarrow$ Supramarginal Gyrus     \\ \hline
			3    &\textbf{Entorhinal Cortex}   $\longleftrightarrow$ Parahippocampal Gyrus        & \textbf{Entorhinal Cortex}   $\longleftrightarrow$   Insula Cortex
			\\ \hline
			4    & \textbf{Entorhinal Cortex}  $\longleftrightarrow$ Frontal Pole & Inferior Temporal Gyrus   $\longleftrightarrow$ Transverse Temporal Cortex
			\\ \hline
			5    & \textbf{Entorhinal Cortex}  $\longleftrightarrow$ Temporal Pole        & Lateral Occipital Cortex   $\longleftrightarrow$ Lateral Orbital Frontal Cortex     \\ \hline \hline
			Rank & NC-ASD LH                                                              & NC-ASD RH                                                            \\ \hline  \hline
			1    & \textbf{Pericalcarine Cortex}   $\longleftrightarrow$ Cuneus Cortex     & \textbf{Pericalcarine Cortex}   $\longleftrightarrow$ Unmeasured Corpus Callosum \\ \hline
			2    &\textbf{Pericalcarine Cortex}   $\longleftrightarrow$  Entorhinal Cortex     & \textbf{Superior Temporal Gyrus}   $\longleftrightarrow$  Inferior Parietal Cortex    \\ \hline
			3    & \textbf{Insula Cortex}   $\longleftrightarrow$ Entorhinal Cortex   & \textbf{Superior Temporal Gyrus}   $\longleftrightarrow$    Lateral Orbital Frontal Cortex     \\ \hline
			4    & \textbf{Pericalcarine Cortex}    $\longleftrightarrow$  Lingual Gyrus & \textbf{Transverse Temporal Cortex}  $\longleftrightarrow$  Pars Orbitalis        \\ \hline
			5    & \textbf{Pericalcarine Cortex}   $\longleftrightarrow$ \textbf{Insula Cortex}        & \textbf{Pericalcarine Cortex} $\longleftrightarrow$ \textbf{Transverse Temporal cortex}       \\ \hline \hline
	\end{tabular}}
	
	\caption{\emph{Top 5 discriminative connections discovered for each population pairs.}}
	\label{tab:3}
\end{table}

% %% ***************************************************************************** %%
\subsection{Biomarker Discovery for  Alzheimer's Disease and Autism Spectrum Disorder.}
\begin{figure}[H]
	\centering
	\includegraphics[width=14cm]{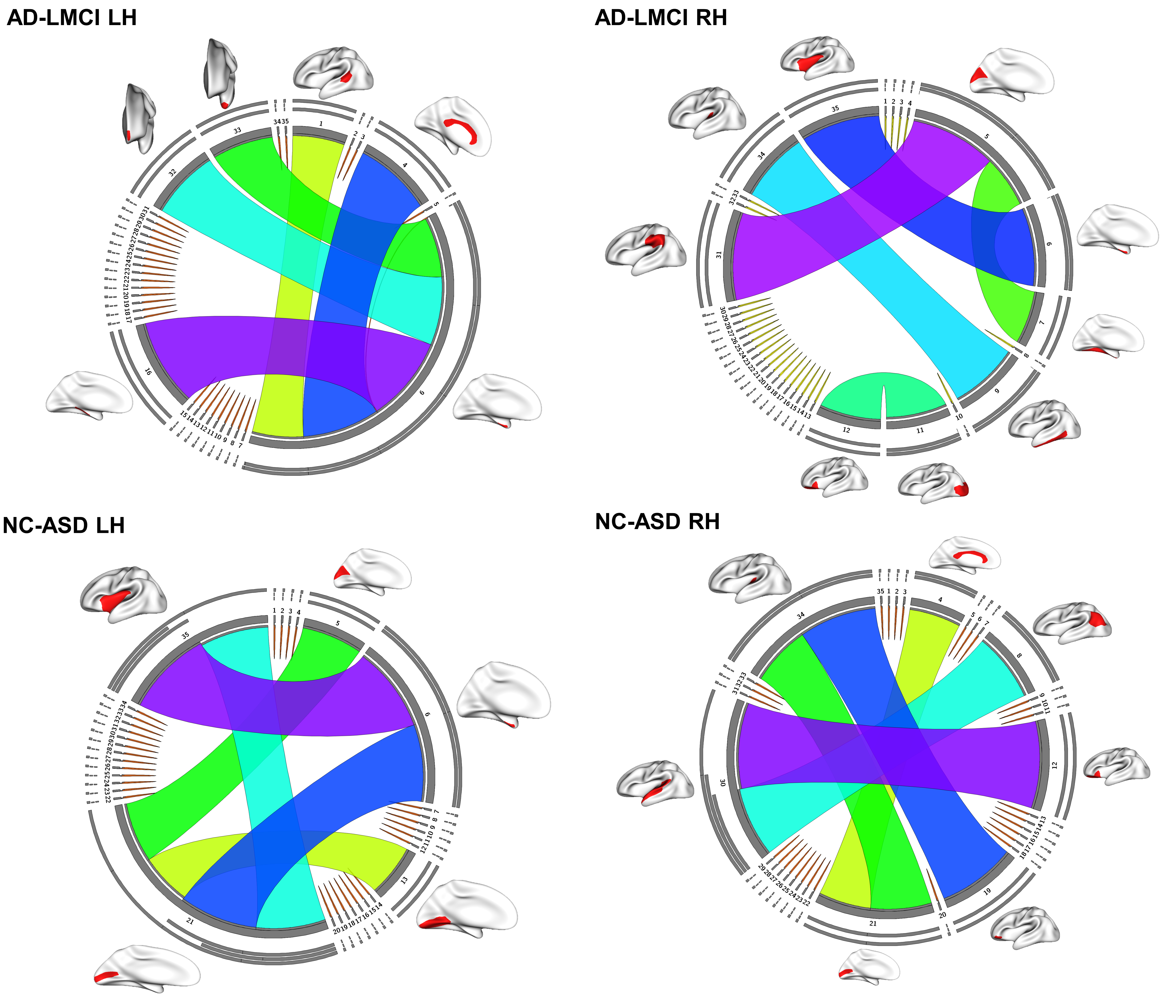}
	\caption{\emph{Circular graphs illustrating the top 5 most discriminative connections that differentiate between two brain states in each pair of groups AD-LMCI and NC-ASD, respectively, for both right and left hemispheres.}
	}
	\label{fig:brains}
\end{figure}

% %% ***************************************************************************** %%
 Given a particular brain disorder, we further investigated whether the connectional features of the learned disordered CBT, with the largest deviations from the learned healthy CBT, present a connectional fingerprint of the disorder of interest. Specifically, we tested if a naive comparison between disordered and healthy population templates provides a meaningful description of how the multi-view connectional aspect of a brain is altered by a particular disorder. Particularly in this study, we leveraged the CBTs generated by MGN-Net to recognize and decipher the connectional morphological alterations of brain regions fingerprinting Alzheimer's disease and the autism spectrum disorder by selecting the top 5 connections with the highest discriminativeness scores for AD-LMCI and NC-ASD populations, respectively (\textbf{Fig.}~\ref{fig:brains}).

\emph{Discovered Alzheimer's disease connectional fingerprint}. We discovered that the most pathologically altered connection between brain regions that differentiates AD patients from LMCI patients are mostly clustered around the entorhinal cortex (EC) and cuneus cortex (CC) (\textbf{Table}~\ref{tab:3}). EC is part of the hippocampal memory system and plays an essential role in memory functions such as memory formation, memory optimization in sleep, and memory consolidation. Therefore, atrophy in EC is likely to be the cause of a significant decline in memory for AD patients when compared to mild symptoms of LMCI \citep{ECAD1,ECAD2}. Moreover, the deviation of CC which is involved in response inhibition \citep{Inhibitory} and generating finger movements based on gaze position \citep{HandPos} might explain the AD strong effect on tasks requiring controlled inhibition processes and motor skills. Insights derived from a simple template comparison also align with existing morphological clinical findings about the demented brain. For example, several studies show that morphological atrophy in the entorhinal cortex is the primary biomarker for the conversion of MCI to AD \citep{EC1,EC2,EC3}. Besides, changes in the volume and the cortical thickness of the CC is also extensively reported as an accurate indicator for the conversion of MCI to AD \citep{CC1,CC2}.

\emph{Discovered autism spectrum disorder connectional fingerprint.} For the large-scale NC-ASD population, MGN-Net identifies the superior temporal gyrus (STG), transverse temporal gyrus (TTG, or Heschl's gyrus), the insular cortex (IC), and the pericalcarine cortex (PC) as top connectional morphological biomarkers of ASD (\textbf{Table}~\ref{tab:3}). STG plays an important role \citep{STGFunc} in auditory, phonetic processing, and social cognition thus, it can explain the receptive language ability deficits which is one of the core features of autism \citep{STGfuncInASD}. Alteration of TTG which is the first cortical structure that processes the auditory information can be the cause of impaired or delayed language abilities of children with autism \citep{TTGASDLang1,TTGASDLang2}. Furthermore, it is extensively reported that the atrophy in the IC which governs the processing of empathy \citep{Empathy}, norm violations \citep{NormVio}, and emotional \citep{EmotionalPro} in the ASD population might be the cause of abnormalities in emotional and affective functions \citep{ICasd}. The alteration of PC which is part of the human visual cortex is likely to be linked to visual symptoms of ASD such as gaze aversion \citep{Gaze}, intense light sensitivity \citep{LightSens}, and disorganized processing of face stimuli \citep{Faces}. Again, from a brain morphology perspective, our connectional ASD blueprint resonates with the existing findings on morphological abnormalities of the autistic brain. For instance, multiple studies showed that a decrease of both white and gray matter volume particularly in the STG \citep{STG} elevated white matter volumes in the TTG \citep{TTG}, significant volumetric increase in IC \citep{IC}, and thicker PC \citep{PC} as common morphological traits that disentangle subjects with autism from healthy subjects.

%\textbf{Ethical approval.}{ This paper does not contain any research conducted by any of the contributors to human respondents. No sources of financing to disclose.}

% %% ***************************************************************************** %%
\section{Discussion and Conclusion}
% %% ***************************************************************************** %%

In this work, we proposed the MGN-Net to generate a normalized connectional template that fingerprints a population of multi-view networks in an end-to-end manner. MGN-Net is topology-aware thanks to its graph neural network layers and novel loss function that preserves domain-specific topological patterns during the integration process. Moreover, MGN-Net is extremely flexible as it is composed of customizable modules that can be useful to the wide community focused on multi-view network integration or normalization stretching from systems biology to social networks. In general, these features of MGN-Net contrast with other network integration methods since they require a set of strong assumptions such as assuming linear relationships among samples, and neglecting distinctive topological characteristics of complex networks.  

From a network neuroscience perspective, we demonstrated that our MGN-Net consistently and significantly outperforms traditional integration methods by generating well-centered, discriminative, and topologically sound connectional templates. Together, our work shows that normalization and integration of multi-view biological graphs can lead to valuable insights by discovering connectional biomarkers that disentangle the typical from the atypical connectivity variability. For example, our connectional brain templates that fingerprint the population of multi-view brain networks derived from T1-weighted MRI scans have revealed a set of biomarkers for both Alzheimer's diseases and the autism spectrum disorder.

Furthermore, conventional methods cannot easily be adapted to more sophisticated network structures such as networks with dynamic connectivity while our flexible architecture needs only small tweaks to operate on any type of network.  In fact, our MGN-Net is powerfully generic and easily adaptable in design to different graph-based problems. For instance, geometric recurrent neural networks (RNNs) based on graph convolutional operation can be used to fuse dynamic brain networks derived from MRI measurements acquired at different time points to reveal the trajectory of neurological diseases \citep{ezzine2019learning,ghribi2021multi,gurler2020foreseeing,nebli2020deep}. Similar to other methods, MGN-Net assumes that all network views contain the same number of nodes. As MGN-Net is rooted in the powerful and adaptable deep learning mindset, it can be naturally extended to handling non-isomorphic graphs with varying numbers of nodes and local topologies by designing graph up-sampling or down-sampling layers. However, the design and physical interpretation of such up-sampling or down-sampling extensions require domain knowledge and may vary across applications. Therefore, in our future work, we intend to devise a build-in extension to MGN-Net that is agnostic to the number of nodes to overcome this limitation. Furthermore, generated connectional templates strongly rely on the selection of the node embedding relationship function. For our case we used the absolute difference, however, this can be also learned to better model the complex and heterogeneous interactions between graph nodes.

We generated representative connectional templates based on a single population for comparative studies. However, MGN-Net can be further enhanced by adding an \emph{auxiliary classification component} right after the graph convolutional layers to create \emph{targeted} connectional templates for tasks that primarily aim to disentangle two specific groups. For example, gender differences in cortical morphological networks can be studied using tailored \emph{targeted} connectional templates method. However, we anticipate that the classification component can force the model to enhance differences between group connectional templates to boost classification results, thus the learned templates can be useful to study specific population pairs.

Like other deep learning models, interpretation of GNNs remains a formidable challenge due to their black-box design.  A new line of research studies this notorious problem. For instance, \citep{GraphLIME} proposed the GraphLIME that is a generic GNN-model explanation framework that interprets a node by generating a model from its $N$-hop neighbors and computes the $k$ most representative features as the explanation using a kernel-based feature selection algorithm. Another work \citep{GNNExplainer} proposed a model-agnostic approach called GNNExplainer for explaining any graph-based machine learning task. GNNExplainer identifies a compact subgraph (with a constraint size) $G_s$ that minimizes the uncertainty of a trained model on a given task when its GNN computation is limited to $G_s$. In our future work, we will explore the potential of these methods to improve the interpretability of our MGN-Net architecture. Their explanations may allow us to identify and fix some of the systematic errors made by MGN-Net.  More importantly, their feature attribution capabilities can be leveraged to guide our integration process to further ensure that the important connectional traits are preserved during the multigraph integration task. For example, we can first train an auxiliary GNN classifier to differentiate several brain network populations, then GNNExplainer can be used to generate a subgraph pattern shared between instances that belong to the same populations. Finally, this graph pattern can be used to assign different importance weights to nodes in the MGN-Net's loss function.

In contrast to other baseline network integration tools, one of the major advantages of our method is that it can be trained collaboratively by different parties without sharing private data based on innovative deep learning strategies such as federated learning \citep{federated} and split neural networks \citep{split}. Furthermore, typically, biomedical data analysis tasks are aggravated by the limited availability of the training data. However, MGN-Net can be trained using transfer learning to achieve good accuracy with limited data. We anticipate that MGN-Net will accelerate the modeling of biological networks and help to ultimately understand the complex dynamics of biological phenomena.

\section*{Precomputation and MGN-Net Infrastructure.} Our publicly available implementation takes a set of stacked adjacency matrices (tensors) as input, however, for efficient graph convolutional layer computation, a couple of preprocessing steps are executed such as casting tensors into different data structures such as edge, feature, and node embedding lists. These precomputation steps run on CPUs and the time and space complexity of these steps depend on the number of subjects, graph nodes as well as the number of views. The main MGN-Net computations including graph convolution operations, pair-wise node embedding relationship computation, and backpropagation are performed on GPUs. MGN-Net training and testing were conducted on a machine with Tesla V100 32GB GPU and  Intel Xeon e5-2698 v4 2.2 GHz (20-core) CPU. As for every deep learning architecture, backpropagation is the most computationally demanding step for MGN-Net as well.  Particularly, our random subset size hyperparameter is the most important factor affecting this computational cost followed by the number of samples, nodes, and views. 

\section*{Data Availability}
The data that support the findings of this study are publicly available from ADNI data (\url{http://adni.loni.usc.edu/}). For reproducibility and comparability, the authors will make available upon request all morphological networks generated based on the four cortical attributes (maximum principal curvature, cortical thickness, sulcal depth, and average curvature) for the 77 subjects (41 AD and 36 LMCI) following the approval by ADNI Consortium. Our large-scale dataset is also available from the public ABIDE initiative (\url{http://fcon\_1000.projects.nitrc.org/indi/abide/}). Following the approval by the ABIDE initiative, all morphological networks generated from the six cortical attributes (cortical surface area and minimum principle area in addition to 4 aforementioned measures) for the 310 subjects (155 NC and 155 ASD) are also accessible from the authors upon request.

\section*{Code Availability}

An open-source Python implementation of MGN-Net is available on GitHub at \url{https://github.com /basiralab/MGN-Net}. The release includes a tutorial, notes regarding Python packages, which need to be installed and, connectional brain templates (CBTs) learned from our four datasets by MGN-Net. Users can directly run MGN-Net either on simulated or externally supplied datasets. Information regarding dataset format can be also found in the same repository. Hyperparameters for the model and training routine can be easily tuned by editing a configuration file.

%\section*{Author Contributions}
%B.G. and I.R. designed the overall method and study and B.G.  developed and implemented the method and software and performed data analysis with input from I.R. B.G. interpreted the results with input from I.R. Both B.G. and I.R. wrote and revised the manuscript. All authors read and approved the final version of the manuscript.

\section*{Competing Interests}
The authors declare no competing interests.

\section{Acknowledgements}

This work was funded by generous grants from the European H2020 Marie Sklodowska-Curie action (grant no. 101003403, \url{http://basira-lab.com/normnets/}) to I.R. and the Scientific and Technological Research Council of Turkey to I.R. under the TUBITAK 2232 Fellowship for Outstanding Researchers (no. 118C288, \url{http://basira-lab.com/reprime/}). However, all scientific contributions made in this project are owned and approved solely by the authors. B.G. is partially supported by the same European H2020 Marie Sklodowska-Curie action.

Data collection and sharing for this project was funded by the Alzheimer's Disease Neuroimaging Initiative (ADNI) (National Institutes of Health Grant U01 AG024904) and DOD ADNI (Department of Defense award number W81XWH-12-2-0012). ADNI is funded by the National Institute on Aging, the National Institute of Biomedical Imaging and Bioengineering, and through generous contributions from the following: AbbVie, Alzheimer's Association; Alzheimer's Drug Discovery Foundation; Araclon Biotech; BioClinica, Inc.; Biogen; Bristol-Myers Squibb Company; CereSpir, Inc.; Cogstate; Eisai Inc.; Elan Pharmaceuticals, Inc.; Eli Lilly and Company; EuroImmun; F. Hoffmann-La Roche Ltd and its affiliated company Genentech, Inc.; Fujirebio; GE Healthcare; IXICO Ltd.; Janssen Alzheimer Immunotherapy Research \& Development, LLC.; Johnson \& Johnson Pharmaceutical Research \& Development LLC.; Lumosity; Lundbeck; Merck \& Co., Inc.; Meso Scale Diagnostics, LLC.; NeuroRx Research; Neurotrack Technologies; Novartis Pharmaceuticals Corporation; Pfizer Inc.; Piramal Imaging; Servier; Takeda Pharmaceutical Company; and Transition Therapeutics. The Canadian Institutes of Health Research is providing funds to support ADNI clinical sites in Canada. Private sector contributions are facilitated by the Foundation for the National Institutes of Health (www.fnih.org). The grantee organization is the Northern California Institute for Research and Education, and the study is coordinated by the Alzheimer's Therapeutic Research Institute at the University of Southern California. ADNI data are disseminated by the Laboratory for Neuro-Imaging at the University of Southern California.

%%%%%%%%%%%%%%%%%%%%%%%%%%%%%%%%%%%%%%%%%%%%%%%%%%%%%%%%%%%%%%%%%%%%%%%%%%%%%%%%%%%%%%%%%%%%%%%%%%%%%%%%%%%%
\newpage
\bibliography{Biblio3}

\begin{thebibliography}{76}
\expandafter\ifx\csname natexlab\endcsname\relax\def\natexlab#1{#1}\fi
\expandafter\ifx\csname url\endcsname\relax
  \def\url#1{\texttt{#1}}\fi
\expandafter\ifx\csname urlprefix\endcsname\relax\def\urlprefix{URL }\fi
\providecommand{\eprint}[2][]{\url{#2}}
\providecommand{\bibinfo}[2]{#2}
\ifx\xfnm\relax \def\xfnm[#1]{\unskip,\space#1}\fi
%Type = Article
\bibitem[{Adams et~al.(1991)Adams, Kelley, Gocayne, Dubnick, Polymeropoulos,
  Xiao, Merril, Wu, Olde, Moreno et~al.}]{HGP}
\bibinfo{author}{Adams, M.}, \bibinfo{author}{Kelley, J.},
  \bibinfo{author}{Gocayne, J.}, \bibinfo{author}{Dubnick, M.},
  \bibinfo{author}{Polymeropoulos, M.}, \bibinfo{author}{Xiao, H.},
  \bibinfo{author}{Merril, C.}, \bibinfo{author}{Wu, A.},
  \bibinfo{author}{Olde, B.}, \bibinfo{author}{Moreno, R.}, et~al.,
  \bibinfo{year}{1991}.
\newblock \bibinfo{title}{Complementary dna sequencing: expressed sequence tags
  and human genome project}.
\newblock \bibinfo{journal}{Science} \bibinfo{volume}{252},
  \bibinfo{pages}{1651--1656}.
%Type = Article
\bibitem[{Bassett and Sporns(2017)}]{NN}
\bibinfo{author}{Bassett, D.}, \bibinfo{author}{Sporns, O.},
  \bibinfo{year}{2017}.
\newblock \bibinfo{title}{Network neuroscience}.
\newblock \bibinfo{journal}{Nature Neuroscience} \bibinfo{volume}{20},
  \bibinfo{pages}{353--364}.
%Type = Article
\bibitem[{Bhagat et~al.(2011)Bhagat, Cormode and Muthukrishnan}]{graph-stats}
\bibinfo{author}{Bhagat, S.}, \bibinfo{author}{Cormode, G.},
  \bibinfo{author}{Muthukrishnan, S.}, \bibinfo{year}{2011}.
\newblock \bibinfo{title}{Node classification in social networks}.
\newblock \bibinfo{journal}{Computing Research Repository - CORR} .
%Type = Article
\bibitem[{Bigler et~al.(2007)Bigler, Mortensen, Neeley, Ozonoff, Krasny,
  Johnson, Lu, Provencal, McMahon and Lainhart}]{STGfuncInASD}
\bibinfo{author}{Bigler, E.}, \bibinfo{author}{Mortensen, S.},
  \bibinfo{author}{Neeley, E.}, \bibinfo{author}{Ozonoff, S.},
  \bibinfo{author}{Krasny, L.}, \bibinfo{author}{Johnson, M.},
  \bibinfo{author}{Lu, J.}, \bibinfo{author}{Provencal, S.},
  \bibinfo{author}{McMahon, W.}, \bibinfo{author}{Lainhart, J.},
  \bibinfo{year}{2007}.
\newblock \bibinfo{title}{Superior temporal gyrus, language function, and
  autism}.
\newblock \bibinfo{journal}{Developmental neuropsychology}
  \bibinfo{volume}{31}, \bibinfo{pages}{217--38}.
%Type = Article
\bibitem[{Bonilha et~al.(2008)Bonilha, Cendes, Rorden, Eckert, Dalgalarrondo,
  Min and Steiner}]{STG}
\bibinfo{author}{Bonilha, L.}, \bibinfo{author}{Cendes, F.},
  \bibinfo{author}{Rorden, C.}, \bibinfo{author}{Eckert, M.},
  \bibinfo{author}{Dalgalarrondo, P.}, \bibinfo{author}{Min, L.},
  \bibinfo{author}{Steiner, C.}, \bibinfo{year}{2008}.
\newblock \bibinfo{title}{Gray and white matter imbalance - typical structural
  abnormality underlying classic autism?}
\newblock \bibinfo{journal}{Brain \& development} \bibinfo{volume}{30},
  \bibinfo{pages}{396--401}.
%Type = Article
\bibitem[{Bronstein et~al.(2017)Bronstein, Bruna, LeCun, Szlam and
  Vandergheynst}]{bronstein2017geometric}
\bibinfo{author}{Bronstein, M.M.}, \bibinfo{author}{Bruna, J.},
  \bibinfo{author}{LeCun, Y.}, \bibinfo{author}{Szlam, A.},
  \bibinfo{author}{Vandergheynst, P.}, \bibinfo{year}{2017}.
\newblock \bibinfo{title}{Geometric deep learning: going beyond euclidean
  data}.
\newblock \bibinfo{journal}{IEEE Signal Processing Magazine}
  \bibinfo{volume}{34}, \bibinfo{pages}{18--42}.
%Type = Article
\bibitem[{Bullmore and Sporns(2009)}]{CBN}
\bibinfo{author}{Bullmore, E.}, \bibinfo{author}{Sporns, O.},
  \bibinfo{year}{2009}.
\newblock \bibinfo{title}{Complex brain networks: Graph theoretical analysis of
  structural and functional systems}.
\newblock \bibinfo{journal}{Nature reviews. Neuroscience} \bibinfo{volume}{10},
  \bibinfo{pages}{186--98}.
%Type = Book
\bibitem[{Burt(1992)}]{effectiveSizel}
\bibinfo{author}{Burt, R.S.}, \bibinfo{year}{1992}.
\newblock \bibinfo{title}{Structural holes: The social structure of
  competition}.
\newblock \bibinfo{publisher}{Harvard University Press},
  \bibinfo{address}{Cambridge, MA}.
%Type = Article
\bibitem[{Bédard and Sanes(2008)}]{HandPos}
\bibinfo{author}{Bédard, P.}, \bibinfo{author}{Sanes, J.},
  \bibinfo{year}{2008}.
\newblock \bibinfo{title}{Gaze and hand position effects on
  finger-movement-related human brain activation}.
\newblock \bibinfo{journal}{Journal of neurophysiology} \bibinfo{volume}{101},
  \bibinfo{pages}{834--42}.
%Type = Article
\bibitem[{Chang et~al.(2010)Chang, Rieger, Johnson, Berger, Barbaro and
  Knight}]{STGFunc}
\bibinfo{author}{Chang, E.}, \bibinfo{author}{Rieger, J.},
  \bibinfo{author}{Johnson, K.}, \bibinfo{author}{Berger, M.},
  \bibinfo{author}{Barbaro, N.}, \bibinfo{author}{Knight, R.},
  \bibinfo{year}{2010}.
\newblock \bibinfo{title}{Categorical speech representation in human superior
  temporal gyrus}.
\newblock \bibinfo{journal}{Nature neuroscience} \bibinfo{volume}{13},
  \bibinfo{pages}{1428--32}.
%Type = Article
\bibitem[{Crockford et~al.(2005)Crockford, Goodyear, Edwards, Quickfall and
  el~Guebaly}]{Inhibitory}
\bibinfo{author}{Crockford, D.}, \bibinfo{author}{Goodyear, B.},
  \bibinfo{author}{Edwards, J.}, \bibinfo{author}{Quickfall, J.},
  \bibinfo{author}{el~Guebaly, N.}, \bibinfo{year}{2005}.
\newblock \bibinfo{title}{Cue-induced brain activity in pathological gamblers}.
\newblock \bibinfo{journal}{Biological psychiatry} \bibinfo{volume}{58},
  \bibinfo{pages}{787--95}.
%Type = Inproceedings
\bibitem[{Defferrard et~al.(2016)Defferrard, Bresson and
  Vandergheynst}]{FastLocalized}
\bibinfo{author}{Defferrard, M.}, \bibinfo{author}{Bresson, X.},
  \bibinfo{author}{Vandergheynst, P.}, \bibinfo{year}{2016}.
\newblock \bibinfo{title}{Convolutional neural networks on graphs with fast
  localized spectral filtering}, in: \bibinfo{editor}{Lee, D.},
  \bibinfo{editor}{Sugiyama, M.}, \bibinfo{editor}{Luxburg, U.},
  \bibinfo{editor}{Guyon, I.}, \bibinfo{editor}{Garnett, R.} (Eds.),
  \bibinfo{booktitle}{Advances in Neural Information Processing Systems},
  \bibinfo{publisher}{Curran Associates, Inc.}. pp.
  \bibinfo{pages}{3844--3852}.
%Type = Article
\bibitem[{Desikan et~al.(2006)Desikan, Ségonne, Fischl, Quinn, Dickerson,
  Blacker, Buckner, Dale, Maguire, Hyman, Albert and Killiany}]{des-kil}
\bibinfo{author}{Desikan, R.}, \bibinfo{author}{Ségonne, F.},
  \bibinfo{author}{Fischl, B.}, \bibinfo{author}{Quinn, B.},
  \bibinfo{author}{Dickerson, B.}, \bibinfo{author}{Blacker, D.},
  \bibinfo{author}{Buckner, R.}, \bibinfo{author}{Dale, A.},
  \bibinfo{author}{Maguire, R.}, \bibinfo{author}{Hyman, B.},
  \bibinfo{author}{Albert, M.}, \bibinfo{author}{Killiany, R.},
  \bibinfo{year}{2006}.
\newblock \bibinfo{title}{An automated labeling system for subdiving the human
  cerebral cortex on mri scans into gyral based regions of interest}.
\newblock \bibinfo{journal}{NeuroImage} \bibinfo{volume}{31},
  \bibinfo{pages}{968--80}.
%Type = Article
\bibitem[{Devanand et~al.(2007)Devanand, Pradhaban, Liu, Khandji, De~Santi,
  Segal, Rusinek, Pelton, Honig, Mayeux, Stern, Tabert and Leon}]{EC2}
\bibinfo{author}{Devanand, D.}, \bibinfo{author}{Pradhaban, G.},
  \bibinfo{author}{Liu, X.}, \bibinfo{author}{Khandji, A.},
  \bibinfo{author}{De~Santi, S.}, \bibinfo{author}{Segal, S.},
  \bibinfo{author}{Rusinek, H.}, \bibinfo{author}{Pelton, G.},
  \bibinfo{author}{Honig, L.}, \bibinfo{author}{Mayeux, R.},
  \bibinfo{author}{Stern, Y.}, \bibinfo{author}{Tabert, M.},
  \bibinfo{author}{Leon, M.}, \bibinfo{year}{2007}.
\newblock \bibinfo{title}{Hippocampal and entorhinal atrophy in mild cognitive
  impairment - prediction of alzheimer disease}.
\newblock \bibinfo{journal}{Neurology} \bibinfo{volume}{68},
  \bibinfo{pages}{828--36}.
%Type = Article
\bibitem[{Dhifallah and Rekik(2019)}]{netnorm}
\bibinfo{author}{Dhifallah, S.}, \bibinfo{author}{Rekik, I.},
  \bibinfo{year}{2019}.
\newblock \bibinfo{title}{Estimation of connectional brain templates using
  selective multi-view network normalization}.
\newblock \bibinfo{journal}{Medical Image Analysis} \bibinfo{volume}{59},
  \bibinfo{pages}{101567}.
%Type = Article
\bibitem[{{Di Martino} et~al.(2014){Di Martino}, Yan, Li, Denio, Castellanos,
  Alaerts, Anderson, Assaf, Bookheimer, Dapretto, Deen, Delmonte, Dinstein,
  Ertl-Wagner, Fair, Gallagher, Kennedy, Keown, Keysers, Lainhart, Lord, Luna,
  Menon, Minshew, Monk, Mueller, M{\"u}ller, Nebel, Nigg, O'Hearn, Pelphrey,
  Peltier, Rudie, Sunaert, Thioux, Tyszka, Uddin, Verhoeven, Wenderoth,
  Wiggins, Mostofsky and Milham}]{ASDdataset}
\bibinfo{author}{{Di Martino}, A.}, \bibinfo{author}{Yan, C.},
  \bibinfo{author}{Li, Q.}, \bibinfo{author}{Denio, E.},
  \bibinfo{author}{Castellanos, F.}, \bibinfo{author}{Alaerts, K.},
  \bibinfo{author}{Anderson, J.}, \bibinfo{author}{Assaf, M.},
  \bibinfo{author}{Bookheimer, S.}, \bibinfo{author}{Dapretto, M.},
  \bibinfo{author}{Deen, B.}, \bibinfo{author}{Delmonte, S.},
  \bibinfo{author}{Dinstein, I.}, \bibinfo{author}{Ertl-Wagner, B.},
  \bibinfo{author}{Fair, D.}, \bibinfo{author}{Gallagher, L.},
  \bibinfo{author}{Kennedy, D.}, \bibinfo{author}{Keown, C.},
  \bibinfo{author}{Keysers, C.}, \bibinfo{author}{Lainhart, J.},
  \bibinfo{author}{Lord, C.}, \bibinfo{author}{Luna, B.},
  \bibinfo{author}{Menon, V.}, \bibinfo{author}{Minshew, N.},
  \bibinfo{author}{Monk, C.}, \bibinfo{author}{Mueller, S.},
  \bibinfo{author}{M{\"u}ller, R.}, \bibinfo{author}{Nebel, M.},
  \bibinfo{author}{Nigg, J.}, \bibinfo{author}{O'Hearn, K.},
  \bibinfo{author}{Pelphrey, K.}, \bibinfo{author}{Peltier, S.},
  \bibinfo{author}{Rudie, J.}, \bibinfo{author}{Sunaert, S.},
  \bibinfo{author}{Thioux, M.}, \bibinfo{author}{Tyszka, J.},
  \bibinfo{author}{Uddin, L.}, \bibinfo{author}{Verhoeven, J.},
  \bibinfo{author}{Wenderoth, N.}, \bibinfo{author}{Wiggins, J.},
  \bibinfo{author}{Mostofsky, S.}, \bibinfo{author}{Milham, M.},
  \bibinfo{year}{2014}.
\newblock \bibinfo{title}{The autism brain imaging data exchange: Towards a
  large-scale evaluation of the intrinsic brain architecture in autism}.
\newblock \bibinfo{journal}{Molecular Psychiatry} \bibinfo{volume}{19},
  \bibinfo{pages}{659--667}.
%Type = Article
\bibitem[{Essen et~al.(2012)Essen, Ugurbil, Auerbach, Barch, Behrens, Bucholz,
  Chang, Chen, Corbetta, Curtiss, Della~Penna, Feinberg, Glasser, Harel, Heath,
  Larson-Prior, Marcus, Michalareas, Moeller and Yacoub}]{HCP}
\bibinfo{author}{Essen, D.}, \bibinfo{author}{Ugurbil, K.},
  \bibinfo{author}{Auerbach, E.}, \bibinfo{author}{Barch, D.},
  \bibinfo{author}{Behrens, T.}, \bibinfo{author}{Bucholz, R.},
  \bibinfo{author}{Chang, A.}, \bibinfo{author}{Chen, L.},
  \bibinfo{author}{Corbetta, M.}, \bibinfo{author}{Curtiss, S.},
  \bibinfo{author}{Della~Penna, S.}, \bibinfo{author}{Feinberg, D.},
  \bibinfo{author}{Glasser, M.}, \bibinfo{author}{Harel, N.},
  \bibinfo{author}{Heath, A.}, \bibinfo{author}{Larson-Prior, L.},
  \bibinfo{author}{Marcus, D.}, \bibinfo{author}{Michalareas, G.},
  \bibinfo{author}{Moeller, S.}, \bibinfo{author}{Yacoub, E.},
  \bibinfo{year}{2012}.
\newblock \bibinfo{title}{The human connectome project: A data acquisition
  perspective}.
\newblock \bibinfo{journal}{NeuroImage} \bibinfo{volume}{62},
  \bibinfo{pages}{2222--31}.
%Type = Inproceedings
\bibitem[{Ezzine and Rekik(2019)}]{ezzine2019learning}
\bibinfo{author}{Ezzine, B.E.}, \bibinfo{author}{Rekik, I.},
  \bibinfo{year}{2019}.
\newblock \bibinfo{title}{Learning-guided infinite network atlas selection for
  predicting longitudinal brain network evolution from a single observation},
  in: \bibinfo{booktitle}{International Conference on Medical Image Computing
  and Computer-Assisted Intervention}, \bibinfo{organization}{Springer}. pp.
  \bibinfo{pages}{796--805}.
%Type = Inproceedings
\bibitem[{Fey and Lenssen(2019)}]{Fey/Lenssen/2019}
\bibinfo{author}{Fey, M.}, \bibinfo{author}{Lenssen, J.E.},
  \bibinfo{year}{2019}.
\newblock \bibinfo{title}{Fast graph representation learning with {PyTorch
  Geometric}}, in: \bibinfo{booktitle}{ICLR Workshop on Representation Learning
  on Graphs and Manifolds}.
%Type = Article
\bibitem[{Fischl(2012)}]{freesurfer}
\bibinfo{author}{Fischl, B.}, \bibinfo{year}{2012}.
\newblock \bibinfo{title}{Freesurfer}.
\newblock \bibinfo{journal}{NeuroImage} \bibinfo{volume}{62},
  \bibinfo{pages}{774--81}.
%Type = Incollection
\bibitem[{Fout et~al.(2017)Fout, Byrd, Shariat and
  Ben-Hur}]{protein-interface-prediction-usingGDL}
\bibinfo{author}{Fout, A.}, \bibinfo{author}{Byrd, J.},
  \bibinfo{author}{Shariat, B.}, \bibinfo{author}{Ben-Hur, A.},
  \bibinfo{year}{2017}.
\newblock \bibinfo{title}{Protein interface prediction using graph
  convolutional networks}, in: \bibinfo{editor}{Guyon, I.},
  \bibinfo{editor}{Luxburg, U.V.}, \bibinfo{editor}{Bengio, S.},
  \bibinfo{editor}{Wallach, H.}, \bibinfo{editor}{Fergus, R.},
  \bibinfo{editor}{Vishwanathan, S.}, \bibinfo{editor}{Garnett, R.} (Eds.),
  \bibinfo{booktitle}{Advances in Neural Information Processing Systems 30}.
  \bibinfo{publisher}{Curran Associates, Inc.}, pp.
  \bibinfo{pages}{6530--6539}.
%Type = Article
\bibitem[{Gainza et~al.(2020)Gainza, Sverrisson, Monti, Rodolà, Boscaini,
  Bronstein and Correia}]{protein-molecular-surfaces-using-gdl}
\bibinfo{author}{Gainza, P.}, \bibinfo{author}{Sverrisson, F.},
  \bibinfo{author}{Monti, F.}, \bibinfo{author}{Rodolà, E.},
  \bibinfo{author}{Boscaini, D.}, \bibinfo{author}{Bronstein, M.},
  \bibinfo{author}{Correia, B.}, \bibinfo{year}{2020}.
\newblock \bibinfo{title}{Deciphering interaction fingerprints from protein
  molecular surfaces using geometric deep learning}.
\newblock \bibinfo{journal}{Nature Methods} \bibinfo{volume}{17},
  \bibinfo{pages}{1--9}.
%Type = Article
\bibitem[{Ghribi et~al.(2021)Ghribi, Li, Lin, Shen and Rekik}]{ghribi2021multi}
\bibinfo{author}{Ghribi, O.}, \bibinfo{author}{Li, G.}, \bibinfo{author}{Lin,
  W.}, \bibinfo{author}{Shen, D.}, \bibinfo{author}{Rekik, I.},
  \bibinfo{year}{2021}.
\newblock \bibinfo{title}{Multi-regression based supervised sample selection
  for predicting baby connectome evolution trajectory from neonatal timepoint}.
\newblock \bibinfo{journal}{Medical Image Analysis} \bibinfo{volume}{68},
  \bibinfo{pages}{101853}.
%Type = Inproceedings
\bibitem[{Gilmer et~al.(2017)Gilmer, Schoenholz, Riley, Vinyals and
  Dahl}]{MPNN}
\bibinfo{author}{Gilmer, J.}, \bibinfo{author}{Schoenholz, S.S.},
  \bibinfo{author}{Riley, P.F.}, \bibinfo{author}{Vinyals, O.},
  \bibinfo{author}{Dahl, G.E.}, \bibinfo{year}{2017}.
\newblock \bibinfo{title}{Neural message passing for quantum chemistry}, in:
  \bibinfo{booktitle}{Proceedings of the 34th International Conference on
  Machine Learning}, pp. \bibinfo{pages}{1263--1272}.
%Type = Article
\bibitem[{Gleich(2014)}]{BeyondWeb}
\bibinfo{author}{Gleich, D.}, \bibinfo{year}{2014}.
\newblock \bibinfo{title}{Pagerank beyond the web}.
\newblock \bibinfo{journal}{SIAM Review} \bibinfo{volume}{57}.
%Type = Article
\bibitem[{Guimerà et~al.(2005)Guimerà, Mossa, Turtschi and
  Amaral}]{transportation}
\bibinfo{author}{Guimerà, R.}, \bibinfo{author}{Mossa, S.},
  \bibinfo{author}{Turtschi, A.}, \bibinfo{author}{Amaral, L.},
  \bibinfo{year}{2005}.
\newblock \bibinfo{title}{The worldwide air transportation network: Anomalous
  centrality, community structure, and cities' global roles}.
\newblock \bibinfo{journal}{Proceedings of the National Academy of Sciences of
  the United States of America} \bibinfo{volume}{102},
  \bibinfo{pages}{7794--9}.
%Type = Inproceedings
\bibitem[{Gurbuz and Rekik(2020)}]{DGN}
\bibinfo{author}{Gurbuz, M.B.}, \bibinfo{author}{Rekik, I.},
  \bibinfo{year}{2020}.
\newblock \bibinfo{title}{Deep graph normalizer: A geometric deep learning
  approach for estimating connectional brain templates}, in:
  \bibinfo{booktitle}{International Conference on Medical Image Computing and
  Computer-Assisted Intervention}, \bibinfo{organization}{Springer}. pp.
  \bibinfo{pages}{155--165}.
%Type = Article
\bibitem[{G{\"u}rler et~al.(2020)G{\"u}rler, Nebli and
  Rekik}]{gurler2020foreseeing}
\bibinfo{author}{G{\"u}rler, Z.}, \bibinfo{author}{Nebli, A.},
  \bibinfo{author}{Rekik, I.}, \bibinfo{year}{2020}.
\newblock \bibinfo{title}{Foreseeing brain graph evolution over time using deep
  adversarial network normalizer}.
\newblock \bibinfo{journal}{International Workshop on PRedictive Intelligence
  In MEdicine} , \bibinfo{pages}{111--122}.
%Type = Misc
\bibitem[{Huang et~al.(2020)Huang, Yamada, Tian, Singh, Yin and
  Chang}]{GraphLIME}
\bibinfo{author}{Huang, Q.}, \bibinfo{author}{Yamada, M.},
  \bibinfo{author}{Tian, Y.}, \bibinfo{author}{Singh, D.},
  \bibinfo{author}{Yin, D.}, \bibinfo{author}{Chang, Y.}, \bibinfo{year}{2020}.
\newblock \bibinfo{title}{Graphlime: Local interpretable model explanations for
  graph neural networks}.
%Type = Article
\bibitem[{Ideker et~al.(2001)Ideker, Galitski and Hood}]{SystemsBio}
\bibinfo{author}{Ideker, T.}, \bibinfo{author}{Galitski, T.},
  \bibinfo{author}{Hood, L.}, \bibinfo{year}{2001}.
\newblock \bibinfo{title}{A new approach to decoding life: Systems biology}.
\newblock \bibinfo{journal}{Annual review of genomics and human genetics}
  \bibinfo{volume}{2}, \bibinfo{pages}{343--72}.
%Type = Article
\bibitem[{Jones et~al.(2003)Jones, Quigney and Huws}]{LightSens}
\bibinfo{author}{Jones, R.}, \bibinfo{author}{Quigney, C.},
  \bibinfo{author}{Huws, J.}, \bibinfo{year}{2003}.
\newblock \bibinfo{title}{First-hand accounts of sensory perceptual experiences
  in autism: A qualitative analysis. journal of intellectual and developmental
  disability, 28(2), 112-121}.
\newblock \bibinfo{journal}{Journal of Intellectual and Developmental
  Disability} \bibinfo{volume}{28}, \bibinfo{pages}{112--121}.
%Type = Article
\bibitem[{Kanehisa and Goto(2000)}]{KEGG}
\bibinfo{author}{Kanehisa, M.}, \bibinfo{author}{Goto, S.},
  \bibinfo{year}{2000}.
\newblock \bibinfo{title}{Kegg: kyoto encyclopedia of genes and genomes}.
\newblock \bibinfo{journal}{Nucleic acids research} \bibinfo{volume}{28},
  \bibinfo{pages}{27--30}.
%Type = Inproceedings
\bibitem[{Kipf and Welling(2017)}]{kipf2016}
\bibinfo{author}{Kipf, T.N.}, \bibinfo{author}{Welling, M.},
  \bibinfo{year}{2017}.
\newblock \bibinfo{title}{{Semi-Supervised Classification with Graph
  Convolutional Networks}}, in: \bibinfo{booktitle}{Proceedings of the 5th
  International Conference on Learning Representations}.
%Type = Misc
\bibitem[{Konečný et~al.(2015)Konečný, McMahan and Ramage}]{federated}
\bibinfo{author}{Konečný, J.}, \bibinfo{author}{McMahan, B.},
  \bibinfo{author}{Ramage, D.}, \bibinfo{year}{2015}.
\newblock \bibinfo{title}{Federated optimization:distributed optimization
  beyond the datacenter}.
\newblock \eprint{1511.03575}.
%Type = Article
\bibitem[{Krizhevsky et~al.(2012)Krizhevsky, Sutskever and Hinton}]{CNNImage}
\bibinfo{author}{Krizhevsky, A.}, \bibinfo{author}{Sutskever, I.},
  \bibinfo{author}{Hinton, G.}, \bibinfo{year}{2012}.
\newblock \bibinfo{title}{Imagenet classification with deep convolutional
  neural networks}.
\newblock \bibinfo{journal}{Neural Information Processing Systems}
  \bibinfo{volume}{25}.
%Type = Article
\bibitem[{Landa and Garrett-Mayer(2006)}]{TTGASDLang1}
\bibinfo{author}{Landa, R.}, \bibinfo{author}{Garrett-Mayer, E.},
  \bibinfo{year}{2006}.
\newblock \bibinfo{title}{Development in infants with autism spectrum
  disorders: A prospective study}.
\newblock \bibinfo{journal}{Journal of child psychology and psychiatry, and
  allied disciplines} \bibinfo{volume}{47}, \bibinfo{pages}{629--38}.
%Type = Article
\bibitem[{Lee et~al.(2008)Lee, Park, Kay, Christakis, Oltvai and
  Barabási}]{metabolic}
\bibinfo{author}{Lee, D.S.}, \bibinfo{author}{Park, J.}, \bibinfo{author}{Kay,
  K.}, \bibinfo{author}{Christakis, N.}, \bibinfo{author}{Oltvai, Z.},
  \bibinfo{author}{Barabási, A.L.}, \bibinfo{year}{2008}.
\newblock \bibinfo{title}{The implications of human metabolic network topology
  for disease comorbidity}.
\newblock \bibinfo{journal}{Proceedings of the National Academy of Sciences of
  the United States of America} \bibinfo{volume}{105},
  \bibinfo{pages}{9880--5}.
%Type = Article
\bibitem[{Liben-nowell and Kleinberg(2003)}]{handcrafted}
\bibinfo{author}{Liben-nowell, D.}, \bibinfo{author}{Kleinberg, J.},
  \bibinfo{year}{2003}.
\newblock \bibinfo{title}{The link prediction problem for social networks}.
\newblock \bibinfo{journal}{Journal of the American Society for Information
  Science and Technology} \bibinfo{volume}{58}.
%Type = Article
\bibitem[{Luyster et~al.(2008)Luyster, Kadlec, Carter and
  Tager-Flusberg}]{TTGASDLang2}
\bibinfo{author}{Luyster, R.}, \bibinfo{author}{Kadlec, M.},
  \bibinfo{author}{Carter, A.}, \bibinfo{author}{Tager-Flusberg, H.},
  \bibinfo{year}{2008}.
\newblock \bibinfo{title}{Language assessment and development in toddlers with
  autism spectrum disorders}.
\newblock \bibinfo{journal}{Journal of autism and developmental disorders}
  \bibinfo{volume}{38}, \bibinfo{pages}{1426--38}.
%Type = Article
\bibitem[{López et~al.(2014)López, Bruña, Aurtenetxe, Pineda-Pardo, Marcos,
  Arrazola, Reinoso, Pedro, Bajo and Maestú}]{EC1}
\bibinfo{author}{López, M.}, \bibinfo{author}{Bruña, R.},
  \bibinfo{author}{Aurtenetxe, S.}, \bibinfo{author}{Pineda-Pardo, J.},
  \bibinfo{author}{Marcos, A.}, \bibinfo{author}{Arrazola, J.},
  \bibinfo{author}{Reinoso, A.}, \bibinfo{author}{Pedro, M.},
  \bibinfo{author}{Bajo, R.}, \bibinfo{author}{Maestú, F.},
  \bibinfo{year}{2014}.
\newblock \bibinfo{title}{Alpha-band hypersynchronization in progressive mild
  cognitive impairment: A magnetoencephalography study}.
\newblock \bibinfo{journal}{Journal of Neuroscience} \bibinfo{volume}{34},
  \bibinfo{pages}{14551--14559}.
%Type = Article
\bibitem[{Mahjoub et~al.(2018)Mahjoub, Mahjoub and Rekik}]{AD-LMCI}
\bibinfo{author}{Mahjoub, I.}, \bibinfo{author}{Mahjoub, M.},
  \bibinfo{author}{Rekik, I.}, \bibinfo{year}{2018}.
\newblock \bibinfo{title}{Brain multiplexes reveal morphological connectional
  biomarkers fingerprinting late brain dementia states}.
\newblock \bibinfo{journal}{Scientific Reports} \bibinfo{volume}{8}.
%Type = Article
\bibitem[{Mirenda et~al.(1983)Mirenda, Donnellan and Yoder}]{Gaze}
\bibinfo{author}{Mirenda, P.}, \bibinfo{author}{Donnellan, A.M.},
  \bibinfo{author}{Yoder, D.E.}, \bibinfo{year}{1983}.
\newblock \bibinfo{title}{Gaze behavior: A new look at an old problem}.
\newblock \bibinfo{journal}{Journal of Autism and Developmental Disorders}
  \bibinfo{volume}{13}, \bibinfo{pages}{397--409}.
%Type = Article
\bibitem[{Mueller et~al.(2005)Mueller, Weiner, Thal, Petersen, Jack, Jagust,
  Trojanowski, Toga and Beckett}]{ADdataset}
\bibinfo{author}{Mueller, S.}, \bibinfo{author}{Weiner, M.},
  \bibinfo{author}{Thal, L.}, \bibinfo{author}{Petersen, R.},
  \bibinfo{author}{Jack, C.}, \bibinfo{author}{Jagust, W.},
  \bibinfo{author}{Trojanowski, J.}, \bibinfo{author}{Toga, A.},
  \bibinfo{author}{Beckett, L.}, \bibinfo{year}{2005}.
\newblock \bibinfo{title}{The alzheimer's disease neuroimaging initiative}.
\newblock \bibinfo{journal}{Neuroimaging clinics of North America}
  \bibinfo{volume}{15}, \bibinfo{pages}{869--77, xi}.
%Type = Article
\bibitem[{Nebli et~al.(2020)Nebli, Kaplan and Rekik}]{nebli2020deep}
\bibinfo{author}{Nebli, A.}, \bibinfo{author}{Kaplan, U.A.},
  \bibinfo{author}{Rekik, I.}, \bibinfo{year}{2020}.
\newblock \bibinfo{title}{Deep evographnet architecture for time-dependent
  brain graph data synthesis from a single timepoint}.
\newblock \bibinfo{journal}{International Workshop on PRedictive Intelligence
  In MEdicine} , \bibinfo{pages}{144--155}.
%Type = Article
\bibitem[{Niskanen et~al.(2011)Niskanen, Könönen, Määttä, Hallikainen,
  Kivipelto, Casarotto, Massimini, Vanninen, Mervaala, Karhu and
  Soininen}]{CC2}
\bibinfo{author}{Niskanen, E.}, \bibinfo{author}{Könönen, M.},
  \bibinfo{author}{Määttä, S.}, \bibinfo{author}{Hallikainen, M.},
  \bibinfo{author}{Kivipelto, M.}, \bibinfo{author}{Casarotto, S.},
  \bibinfo{author}{Massimini, M.}, \bibinfo{author}{Vanninen, R.},
  \bibinfo{author}{Mervaala, E.}, \bibinfo{author}{Karhu, J.},
  \bibinfo{author}{Soininen, H.}, \bibinfo{year}{2011}.
\newblock \bibinfo{title}{New insights into alzheimer's disease progression: A
  combined tms and structural mri study}.
\newblock \bibinfo{journal}{PloS one} \bibinfo{volume}{6},
  \bibinfo{pages}{e26113}.
%Type = Article
\bibitem[{Onnela et~al.(2005)Onnela, Saramäki, Kertész and
  Kaski}]{clustering}
\bibinfo{author}{Onnela, J.P.}, \bibinfo{author}{Saramäki, J.},
  \bibinfo{author}{Kertész, J.}, \bibinfo{author}{Kaski, K.},
  \bibinfo{year}{2005}.
\newblock \bibinfo{title}{Intensity and coherence of motifs in weighted complex
  networks}.
\newblock \bibinfo{journal}{Physical review. E, Statistical, nonlinear, and
  soft matter physics} \bibinfo{volume}{71}, \bibinfo{pages}{065103}.
%Type = Inproceedings
\bibitem[{Page et~al.()Page, Brin, Motwani and Winograd}]{pagerank}
\bibinfo{author}{Page, L.}, \bibinfo{author}{Brin, S.},
  \bibinfo{author}{Motwani, R.}, \bibinfo{author}{Winograd, T.}, .
\newblock \bibinfo{title}{The pagerank citation ranking: Bringing order to the
  web}, in: \bibinfo{booktitle}{Proceedings of the 7th International World Wide
  Web Conference}, \bibinfo{address}{Brisbane, Australia}. pp.
  \bibinfo{pages}{161--172}.
%Type = Article
\bibitem[{Parisot et~al.(2018)Parisot, Ktena, Ferrante, Lee, Guerrero, Glocker
  and Rueckert}]{ASD-and-AD}
\bibinfo{author}{Parisot, S.}, \bibinfo{author}{Ktena, S.I.},
  \bibinfo{author}{Ferrante, E.}, \bibinfo{author}{Lee, M.},
  \bibinfo{author}{Guerrero, R.}, \bibinfo{author}{Glocker, B.},
  \bibinfo{author}{Rueckert, D.}, \bibinfo{year}{2018}.
\newblock \bibinfo{title}{Disease prediction using graph convolutional
  networks: Application to autism spectrum disorder and alzheimer’s disease}.
\newblock \bibinfo{journal}{Medical Image Analysis} \bibinfo{volume}{48}.
%Type = Incollection
\bibitem[{Pearl(1988)}]{Message}
\bibinfo{author}{Pearl, J.}, \bibinfo{year}{1988}.
\newblock in: \bibinfo{editor}{Pearl, J.} (Ed.),
  \bibinfo{booktitle}{Probabilistic Reasoning in Intelligent Systems}.
  \bibinfo{publisher}{Morgan Kaufmann}, \bibinfo{address}{San Francisco (CA)}.
%Type = Article
\bibitem[{Pedregosa et~al.(2011)Pedregosa, Varoquaux, Gramfort, Michel,
  Thirion, Grisel, Blondel, Prettenhofer, Weiss, Dubourg, Vanderplas, Passos,
  Cournapeau, Brucher, Perrot and Duchesnay}]{scikit-learn}
\bibinfo{author}{Pedregosa, F.}, \bibinfo{author}{Varoquaux, G.},
  \bibinfo{author}{Gramfort, A.}, \bibinfo{author}{Michel, V.},
  \bibinfo{author}{Thirion, B.}, \bibinfo{author}{Grisel, O.},
  \bibinfo{author}{Blondel, M.}, \bibinfo{author}{Prettenhofer, P.},
  \bibinfo{author}{Weiss, R.}, \bibinfo{author}{Dubourg, V.},
  \bibinfo{author}{Vanderplas, J.}, \bibinfo{author}{Passos, A.},
  \bibinfo{author}{Cournapeau, D.}, \bibinfo{author}{Brucher, M.},
  \bibinfo{author}{Perrot, M.}, \bibinfo{author}{Duchesnay, E.},
  \bibinfo{year}{2011}.
\newblock \bibinfo{title}{Scikit-learn: Machine learning in {P}ython}.
\newblock \bibinfo{journal}{Journal of Machine Learning Research}
  \bibinfo{volume}{12}, \bibinfo{pages}{2825--2830}.
%Type = Article
\bibitem[{Pelphrey et~al.(2002)Pelphrey, Sasson, Reznick, Paul, Goldman and
  Piven}]{Faces}
\bibinfo{author}{Pelphrey, K.}, \bibinfo{author}{Sasson, N.},
  \bibinfo{author}{Reznick, J.}, \bibinfo{author}{Paul, G.},
  \bibinfo{author}{Goldman, B.}, \bibinfo{author}{Piven, J.},
  \bibinfo{year}{2002}.
\newblock \bibinfo{title}{Visual scanning of faces in autism}.
\newblock \bibinfo{journal}{Journal of autism and developmental disorders}
  \bibinfo{volume}{32}, \bibinfo{pages}{249--61}.
%Type = Article
\bibitem[{Phan et~al.(2002)Phan, Wager, Taylor and Liberzon}]{EmotionalPro}
\bibinfo{author}{Phan, K.L.}, \bibinfo{author}{Wager, T.D.},
  \bibinfo{author}{Taylor, S.F.}, \bibinfo{author}{Liberzon, I.},
  \bibinfo{year}{2002}.
\newblock \bibinfo{title}{Functional neuroanatomy of emotion: A meta-analysis
  of emotion activation studies in pet and fmri}.
\newblock \bibinfo{journal}{NeuroImage} \bibinfo{volume}{16},
  \bibinfo{pages}{331--348}.
%Type = Misc
\bibitem[{Rhee et~al.(2018)Rhee, Seo and
  Kim}]{Breast-Cancer-Subtype-Classification}
\bibinfo{author}{Rhee, S.}, \bibinfo{author}{Seo, S.}, \bibinfo{author}{Kim,
  S.}, \bibinfo{year}{2018}.
\newblock \bibinfo{title}{Hybrid approach of relation network and localized
  graph convolutional filtering for breast cancer subtype classification}.
\newblock \eprint{1711.05859}.
%Type = Article
\bibitem[{Safari et~al.(2014)Safari, Taghizadeh, Rezaei~tavirani, Goliaei and
  Peyvandi}]{PPI}
\bibinfo{author}{Safari, N.}, \bibinfo{author}{Taghizadeh, M.},
  \bibinfo{author}{Rezaei~tavirani, M.}, \bibinfo{author}{Goliaei, B.},
  \bibinfo{author}{Peyvandi, A.}, \bibinfo{year}{2014}.
\newblock \bibinfo{title}{Protein-protein interaction networks (ppi) and
  complex diseases}.
\newblock \bibinfo{journal}{Gastroenterology and hepatology from bed to bench}
  \bibinfo{volume}{7}, \bibinfo{pages}{17--31}.
%Type = Article
\bibitem[{Salathé et~al.(2010)Salathé, Kazandjieva, Lee, Levis, Feldman and
  Jones}]{contactNetwork}
\bibinfo{author}{Salathé, M.}, \bibinfo{author}{Kazandjieva, M.},
  \bibinfo{author}{Lee, J.W.}, \bibinfo{author}{Levis, P.},
  \bibinfo{author}{Feldman, M.}, \bibinfo{author}{Jones, J.},
  \bibinfo{year}{2010}.
\newblock \bibinfo{title}{A high-resolution human contact network for
  infectious disease transmission}.
\newblock \bibinfo{journal}{Proceedings of the National Academy of Sciences of
  the United States of America} \bibinfo{volume}{107},
  \bibinfo{pages}{22020--5}.
%Type = Article
\bibitem[{Sanfey et~al.(2003)Sanfey, Rilling, Aronson, Nystrom and
  Cohen}]{NormVio}
\bibinfo{author}{Sanfey, A.G.}, \bibinfo{author}{Rilling, J.K.},
  \bibinfo{author}{Aronson, J.A.}, \bibinfo{author}{Nystrom, L.},
  \bibinfo{author}{Cohen, J.D.}, \bibinfo{year}{2003}.
\newblock \bibinfo{title}{The neural basis of economic decision-making in the
  ultimatum game.}
\newblock \bibinfo{journal}{Science} \bibinfo{volume}{300 5626},
  \bibinfo{pages}{1755--8}.
%Type = Inproceedings
\bibitem[{Simonovsky and Komodakis(2017)}]{edgeConditioned}
\bibinfo{author}{Simonovsky, M.}, \bibinfo{author}{Komodakis, N.},
  \bibinfo{year}{2017}.
\newblock \bibinfo{title}{Dynamic edge-conditioned filters in convolutional
  neural networks on graphs}, in: \bibinfo{booktitle}{CVPR}.
%Type = Article
\bibitem[{Singer(2006)}]{Empathy}
\bibinfo{author}{Singer, T.}, \bibinfo{year}{2006}.
\newblock \bibinfo{title}{The neuronal basis and ontogeny of empathy and mind
  reading: Review of literature and implications for future research}.
\newblock \bibinfo{journal}{Neuroscience and biobehavioral reviews}
  \bibinfo{volume}{30}, \bibinfo{pages}{855--63}.
%Type = Article
\bibitem[{Soussia and Rekik(2017)}]{soussia2017high}
\bibinfo{author}{Soussia, M.}, \bibinfo{author}{Rekik, I.},
  \bibinfo{year}{2017}.
\newblock \bibinfo{title}{High-order connectomic manifold learning for autistic
  brain state identification}.
\newblock \bibinfo{journal}{International Workshop on Connectomics in
  Neuroimaging} , \bibinfo{pages}{51--59}.
%Type = Article
\bibitem[{Soussia and Rekik(2018)}]{Soussia:2018b}
\bibinfo{author}{Soussia, M.}, \bibinfo{author}{Rekik, I.},
  \bibinfo{year}{2018}.
\newblock \bibinfo{title}{Unsupervised manifold learning using high-order
  morphological brain networks derived from {T1-w MRI} for autism diagnosis}.
\newblock \bibinfo{journal}{Frontiers in Neuroinformatics}
  \bibinfo{volume}{12}.
%Type = Article
\bibitem[{Van~Essen and Glasser(2016)}]{van2016human}
\bibinfo{author}{Van~Essen, D.C.}, \bibinfo{author}{Glasser, M.F.},
  \bibinfo{year}{2016}.
\newblock \bibinfo{title}{The human connectome project: progress and
  prospects}.
\newblock \bibinfo{journal}{Cerebrum: the Dana forum on brain science}
  \bibinfo{volume}{2016}.
%Type = Article
\bibitem[{Van~Hoesen et~al.(1991)Van~Hoesen, Hyman and Damasio}]{ECAD2}
\bibinfo{author}{Van~Hoesen, G.}, \bibinfo{author}{Hyman, B.},
  \bibinfo{author}{Damasio, A.}, \bibinfo{year}{1991}.
\newblock \bibinfo{title}{Entorhinal cortex pathology in alzheimer's disease}.
\newblock \bibinfo{journal}{Hippocampus} \bibinfo{volume}{1},
  \bibinfo{pages}{1—8}.
%Type = Inproceedings
\bibitem[{Veličković et~al.(2018)Veličković, Cucurull, Casanova, Romero,
  Liò and Bengio}]{GAttention}
\bibinfo{author}{Veličković, P.}, \bibinfo{author}{Cucurull, G.},
  \bibinfo{author}{Casanova, A.}, \bibinfo{author}{Romero, A.},
  \bibinfo{author}{Liò, P.}, \bibinfo{author}{Bengio, Y.},
  \bibinfo{year}{2018}.
\newblock \bibinfo{title}{Graph attention networks}, in:
  \bibinfo{booktitle}{International Conference on Learning Representations}.
%Type = Misc
\bibitem[{Vepakomma et~al.(2018)Vepakomma, Gupta, Swedish and Raskar}]{split}
\bibinfo{author}{Vepakomma, P.}, \bibinfo{author}{Gupta, O.},
  \bibinfo{author}{Swedish, T.}, \bibinfo{author}{Raskar, R.},
  \bibinfo{year}{2018}.
\newblock \bibinfo{title}{Split learning for health: Distributed deep learning
  without sharing raw patient data}.
%Type = Article
\bibitem[{Vishwanathan et~al.(2008)Vishwanathan, Borgwardt, Kondor and
  Schraudolph}]{kernels}
\bibinfo{author}{Vishwanathan, S.}, \bibinfo{author}{Borgwardt, K.},
  \bibinfo{author}{Kondor, R.}, \bibinfo{author}{Schraudolph, N.},
  \bibinfo{year}{2008}.
\newblock \bibinfo{title}{Graph kernels}.
\newblock \bibinfo{journal}{J Mach Learn Res} \bibinfo{volume}{11}.
%Type = Article
\bibitem[{Wang et~al.(2014)Wang, Mezlini, Demir, Fiume, Tu, Brudno, Haibe-Kains
  and Goldenberg}]{SNF}
\bibinfo{author}{Wang, B.}, \bibinfo{author}{Mezlini, A.},
  \bibinfo{author}{Demir, F.}, \bibinfo{author}{Fiume, M.},
  \bibinfo{author}{Tu, Z.}, \bibinfo{author}{Brudno, M.},
  \bibinfo{author}{Haibe-Kains, B.}, \bibinfo{author}{Goldenberg, A.},
  \bibinfo{year}{2014}.
\newblock \bibinfo{title}{Similarity network fusion for aggregating data types
  on a genomic scale}.
\newblock \bibinfo{journal}{Nature methods} \bibinfo{volume}{11}.
%Type = Inbook
\bibitem[{Watts et~al.(2006)Watts, J, Strogatz and H}]{smallworld}
\bibinfo{author}{Watts, D.}, \bibinfo{author}{J, D.},
  \bibinfo{author}{Strogatz}, \bibinfo{author}{H, S.}, \bibinfo{year}{2006}.
\newblock \bibinfo{title}{Collective dynamics of 'small world' networks}.
\newblock pp. \bibinfo{pages}{301--303}.
%Type = Article
\bibitem[{Wee et~al.(2013)Wee, Yap and Shen}]{CC1}
\bibinfo{author}{Wee, C.Y.}, \bibinfo{author}{Yap, P.T.},
  \bibinfo{author}{Shen, D.}, \bibinfo{year}{2013}.
\newblock \bibinfo{title}{Prediction of alzheimer's disease and mild cognitive
  impairment using cortical morphological patterns}.
\newblock \bibinfo{journal}{Human brain mapping} \bibinfo{volume}{34}.
%Type = Article
\bibitem[{Whitwell et~al.(2007)Whitwell, Przybelski, Weigand, Knopman, Boeve,
  Petersen and Jack}]{EC3}
\bibinfo{author}{Whitwell, J.}, \bibinfo{author}{Przybelski, S.},
  \bibinfo{author}{Weigand, S.}, \bibinfo{author}{Knopman, D.},
  \bibinfo{author}{Boeve, B.}, \bibinfo{author}{Petersen, R.},
  \bibinfo{author}{Jack, C.}, \bibinfo{year}{2007}.
\newblock \bibinfo{title}{3d maps from multiple mri illustrate changing atrophy
  patterns as subjects progress from mild cognitive impairment to alzheimer's
  disease}.
\newblock \bibinfo{journal}{Brain : a journal of neurology}
  \bibinfo{volume}{130}, \bibinfo{pages}{1777--86}.
%Type = Article
\bibitem[{Xiao et~al.(2014)Xiao, Qiu, Ke, Xiao, Xiao, Liang, Zou, Huang, Fang,
  Chu, Zhang and Liu}]{TTG}
\bibinfo{author}{Xiao, Z.}, \bibinfo{author}{Qiu, T.}, \bibinfo{author}{Ke,
  X.}, \bibinfo{author}{Xiao, X.}, \bibinfo{author}{Xiao, T.},
  \bibinfo{author}{Liang, F.}, \bibinfo{author}{Zou, B.},
  \bibinfo{author}{Huang, H.}, \bibinfo{author}{Fang, H.},
  \bibinfo{author}{Chu, K.}, \bibinfo{author}{Zhang, J.}, \bibinfo{author}{Liu,
  Y.}, \bibinfo{year}{2014}.
\newblock \bibinfo{title}{Autism spectrum disorder as early neurodevelopmental
  disorder: Evidence from the brain imaging abnormalities in 2–3 years old
  toddlers}.
\newblock \bibinfo{journal}{Journal of autism and developmental disorders}
  \bibinfo{volume}{44}.
%Type = Misc
\bibitem[{Xu et~al.(2019)Xu, Hu, Leskovec and Jegelka}]{xu2019powerful}
\bibinfo{author}{Xu, K.}, \bibinfo{author}{Hu, W.}, \bibinfo{author}{Leskovec,
  J.}, \bibinfo{author}{Jegelka, S.}, \bibinfo{year}{2019}.
\newblock \bibinfo{title}{How powerful are graph neural networks?}
\newblock \eprint{1810.00826}.
%Type = Article
\bibitem[{Yamada et~al.(2016a)Yamada, Itahashi, Nakamura, Watanabe, Kuroda,
  Ohta, Kanai, Kato and Hashimoto}]{ICasd}
\bibinfo{author}{Yamada, T.}, \bibinfo{author}{Itahashi, T.},
  \bibinfo{author}{Nakamura, M.}, \bibinfo{author}{Watanabe, H.},
  \bibinfo{author}{Kuroda, M.}, \bibinfo{author}{Ohta, H.},
  \bibinfo{author}{Kanai, C.}, \bibinfo{author}{Kato, N.},
  \bibinfo{author}{Hashimoto, R.i.}, \bibinfo{year}{2016}a.
\newblock \bibinfo{title}{Altered functional organization within the insular
  cortex in adult males with high- functioning autism spectrum disorder:
  evidence from connectivity-based parcellation}.
\newblock \bibinfo{journal}{Molecular Autism} \bibinfo{volume}{7}.
%Type = Article
\bibitem[{Yamada et~al.(2016b)Yamada, Itahashi, Nakamura, Watanabe, Kuroda,
  Ohta, Kanai, Kato and Hashimoto}]{IC}
\bibinfo{author}{Yamada, T.}, \bibinfo{author}{Itahashi, T.},
  \bibinfo{author}{Nakamura, M.}, \bibinfo{author}{Watanabe, H.},
  \bibinfo{author}{Kuroda, M.}, \bibinfo{author}{Ohta, H.},
  \bibinfo{author}{Kanai, C.}, \bibinfo{author}{Kato, N.},
  \bibinfo{author}{Hashimoto, R.i.}, \bibinfo{year}{2016}b.
\newblock \bibinfo{title}{Altered functional organization within the insular
  cortex in adult males with high- functioning autism spectrum disorder:
  evidence from connectivity-based parcellation}.
\newblock \bibinfo{journal}{Molecular Autism} \bibinfo{volume}{7}.
%Type = Article
\bibitem[{Yassa(2014)}]{ECAD1}
\bibinfo{author}{Yassa, M.}, \bibinfo{year}{2014}.
\newblock \bibinfo{title}{Ground zero in alzheimer's disease}.
\newblock \bibinfo{journal}{Nature neuroscience} \bibinfo{volume}{17},
  \bibinfo{pages}{146--7}.
%Type = Article
\bibitem[{Ying et~al.(2019)Ying, Bourgeois, You, Zitnik and
  Leskovec}]{GNNExplainer}
\bibinfo{author}{Ying, R.}, \bibinfo{author}{Bourgeois, D.},
  \bibinfo{author}{You, J.}, \bibinfo{author}{Zitnik, M.},
  \bibinfo{author}{Leskovec, J.}, \bibinfo{year}{2019}.
\newblock \bibinfo{title}{Gnnexplainer: Generating explanations for graph
  neural networks}.
\newblock \bibinfo{journal}{Advances in neural information processing systems}
  \bibinfo{volume}{32}, \bibinfo{pages}{9240--9251}.
%Type = Article
\bibitem[{Zielinski et~al.(2014)Zielinski, Prigge, Nielsen, Froehlich,
  Abildskov, Anderson, Fletcher, Zygmunt, Travers, Lange, Alexander, Bigler and
  Lainhart}]{PC}
\bibinfo{author}{Zielinski, B.}, \bibinfo{author}{Prigge, M.},
  \bibinfo{author}{Nielsen, J.}, \bibinfo{author}{Froehlich, A.},
  \bibinfo{author}{Abildskov, T.}, \bibinfo{author}{Anderson, J.},
  \bibinfo{author}{Fletcher, P.}, \bibinfo{author}{Zygmunt, K.},
  \bibinfo{author}{Travers, B.}, \bibinfo{author}{Lange, N.},
  \bibinfo{author}{Alexander, A.}, \bibinfo{author}{Bigler, E.},
  \bibinfo{author}{Lainhart, J.}, \bibinfo{year}{2014}.
\newblock \bibinfo{title}{Longitudinal changes in cortical thickness in autism
  and typical development}.
\newblock \bibinfo{journal}{Brain : a journal of neurology}
  \bibinfo{volume}{137}.

\end{thebibliography}
\bibliographystyle{model2-names}
\end{document}